\documentclass[manuscript, screen, nonacm]{acmart}

\usepackage{multirow}
\usepackage{multicol}
\usepackage{arydshln}
\usepackage{graphicx}
\usepackage{balance}
\usepackage{adjustbox}
\usepackage{textcomp}
\usepackage{balance}
\usepackage{xcolor}
\usepackage{soul}
\usepackage{booktabs}
\usepackage{graphicx}
\usepackage[caption=false]{subfig}

\usepackage{comment}
\usepackage{colortbl}

\AtBeginDocument{%
  \providecommand\BibTeX{{%
    \normalfont B\kern-0.5em{\scshape i\kern-0.25em b}\kern-0.8em\TeX}}}

\setcopyright{acmcopyright}
\copyrightyear{2024}
\acmYear{2024}
\acmDOI{XXXXXXX.XXXXXXX}
\def\eg{\emph{e.g., }}

\newcommand{\kpnote}[1]{\noindent \textcolor{blue}{KP: #1 }}

\author{Kaitlynn Taylor Pineda}
\email{kpineda3@jhu.edu}

\affiliation{%
  \institution{Johns Hopkins University}
  \streetaddress{3400 N. Charles St}
  \city{Baltimore}
  \state{Maryland}
  \country{USA}
  \postcode{21218}
}

\author{Amama Mahmood}
\email{amama.mahmood@jhu.edu}
 
\affiliation{%
  \institution{Johns Hopkins University}
  \streetaddress{3400 N. Charles St}
  \city{Baltimore}
  \state{Maryland}
  \country{USA}
  \postcode{21218}
}

\author{Juo-Tung Chen}
\email{jchen396@jhu.edu}

\affiliation{%
  \institution{Johns Hopkins University}
  \streetaddress{3400 N. Charles St}
  \city{Baltimore}
  \state{Maryland}
  \country{USA}
  \postcode{21218}
}
   
\author{Chien-Ming Huang}
\email{cmhuang@cs.jhu.edu}

\affiliation{%
  \institution{Johns Hopkins University}
  \streetaddress{3400 N. Charles St}
  \city{Baltimore}
  \state{Maryland}
  \country{USA}
  \postcode{21218}
}

\renewcommand{\shortauthors}{Pineda, Mahmood, Chen and Huang}

\begin{document}

\title[How People Respond to Small Talk During HRC]{``You Might Like It'': How People Respond to Small Talk During Human-Robot Collaboration}

\begin{teaserfigure}
  \includegraphics[width=\textwidth]{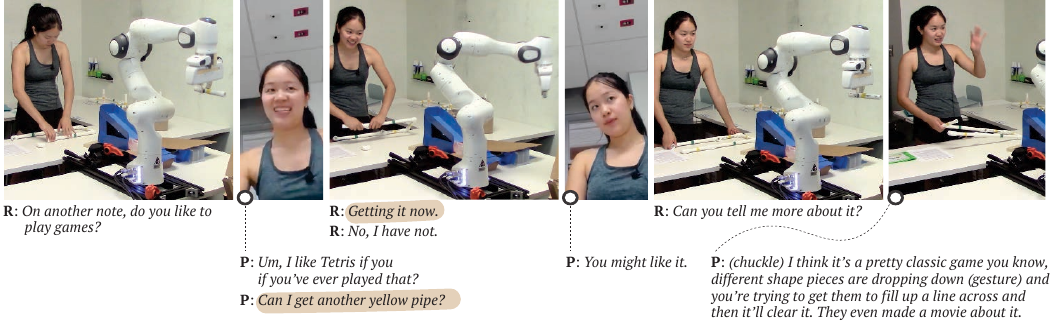}
  \caption{We explore participants’ interactions with a manipulator robot that engages in small-talk during a collaborative assembly task. This figure shows the sequence of some spoken phrases shared by the participant (P1) and robot; task-oriented dialogue is interchanged with small-talk. The task-oriented requests and responses are highlighted. }
  \Description{This figure includes 6 photos, alternating between the participant themselves and the participant and the robot interacting together. Below is text for each image that depicts the conversation that is exchanged between the human and the robot. The conversation is as follows: robot: on another note, do you like to play any games?;
participant:  um, I like Tetris if you if you've ever played that?;
participant: Can I get another yellow pipe?;
robot: Getting it now.;
robot:  No, I have not;
participant: you might like it;
robot:  Can you tell me more about it?;
participant: [chuckle] I think it's a pretty classic game you know you just like the pieces different shape pieces are dropping down and you're trying to get them to fill up a line across and then it'll like clear it. they even made a movie about it.}
  \label{fig:teaser}
\end{teaserfigure}

\begin{abstract}

Social communication between people and social robots has been studied extensively and found to have various notable benefits, including the enhancement of human-robot team cohesion and the development of rapport and trust. However, the potential of social communication between people and non-social robots, such as non-anthropomorphic robot manipulators commonly used in work settings (\eg warehouse and factory), is less explored and not well established. In this work, we investigate people's engagement and attitudes towards a non-anthropomorphic robot manipulator that initiates small talk during a collaborative assembly task and explore how the presence of negative team feedback may affect team dynamics and blame attribution. Through an in-person study with 20 participants, we observed a response rate of $77.60\%$ in response to the robot's small talk attempts. Nine participants continued engaging with the robot by initiating their own questions, indicating sustained interest in the conversation. However, we also found that the first negative feedback decreased the participants' willingness to extend the conversation. We additionally present participants' initial perceptions of small talk for physical robot manipulators and discuss design implications for integrating small talk into non-social robots, along with various aspects of small talk that may influence physical human-robot interactions.

\end{abstract}



\maketitle

\section{Introduction}


Human-robot collaboration research envisions robots as human collaborators, or co-workers, to enhance human work in diverse domains. Take manufacturing as an example: the manufacturing industry is currently undergoing a significant transformation to become more flexible and agile, allowing it to accommodate mass product customization even while product life cycles are continuously decreasing. Moreover, small and medium-sized manufacturing companies face a shortage of skilled production workers. Collaborative robots---robots capable of working alongside people on joint tasks safely and efficiently---promise to fill the needed jobs, enable agile manufacturing, and increase productivity.

Even though these collaborative robots are positioned as co-workers, they are mostly non-anthropomorphic as in the form of robotic arm (manipulator) and are designed to support functional, physical tasks. 
Human collaboration, on the other hand, is more than just performing the joint task but involves off-task social interactions, which are integral to establishing and maintaining user trust and rapport, key ingredients for continued productive collaboration \cite{kopp2021success}. 
While prior research has found that people like to have and expect social interactions with their robot co-workers in manufacturing settings \cite{Sauppé_Mutlu_2015}, the robots studied were the Baxter robot with animated facial expressions and a humanlike form factor. It is unclear if people would engage in meaningful social interactions with non-anthropomorphic robot manipulators.

In this work, we explore the possibility of a non-anthropomorphic robot manipulator not just working with people to complete physical tasks but also engaging in social exchanges, particularly small talk---commonly seen in human interactions for rapport and trust building \cite{Laver1981LinguisticRA}---as human co-workers would.
Specifically, we sought to study whether and how people respond to small talk from a robot manipulator during a collaborative task. 
Moreover, we were interested in understanding how negative performance feedback to the human-robot team might shape the team dynamics and blame attribution; this inquiry is relevant and of interest because of the rising adoption of performance monitoring systems in industrial settings (\eg warehouse). 
Through a study with 20 participants, we found that most participants ($n=18$) interacted socially with the robot (Fig. \ref{fig:teaser}), nine of whom even initiated questions back to the robot. 
The presence of initial negative team feedback significantly decreased participants' average tendency to extend their small talk conversation with the robot. 
We explore the potential for integrating small talk into collaborative robots and discuss various factors 
that could affect the dynamics of human-robot small talk interactions in the future.

\section{Background and Related Work}


We first review prior works that study how to incorporate social behaviors into robots across various domains, then highlight works that discuss how people treat machines similar to other humans before finally introducing background on small talk itself. 

\subsection{Domains of Social Human-Robot Interactions}


Researchers have studied how robots can interact socially with humans in various domains such as for tutoring and educational contexts \cite{belpaeme2018education, salomons2022we},
encouraging exercise \cite{antony2023co}, providing companionship \cite{kidd2006sociable, tsoi2021challenges}, and emotional support \cite{sabelli2011conversational}, and other therapeutic interventions  \cite{feil2009toward, scassellati2018improving}. 
Many of these engagements explore integrating various multi-modal capabilities in robots to promote their engagements with humans \cite{youssef2022survey, huang2013repertoire}. These can consist of 
robot eye gaze behaviors \cite{mutlu2009footing, admoni2017social}, head movements \cite{sidner2006effect, kkedzierski2013emys}, and gestures \cite{meena2012integration} alongside robot speech abilities \cite{striepe2021modeling, sinnema2019attitude}. 
While research has been conducted to analyze the characteristics of conversational interactions with robots \cite{seok2022cultural, laban2020tell}, most of these works utilize social robots as their robotic entity. 
However, the prevalence of non-social robots, such as cobots or other manipulator robots, is much greater than that of social robots themselves, and few efforts have been made to explore further what will happen if we have conversations between humans and non-social robots. 
One group of researchers utilized storyboards to depict cobots exhibiting affective and playful behaviors and fostering a sense of relatedness with the user in a physical, collaborative scenario \cite{chowdhury2021you}. 
Another utilized a Wizard-of-Oz labeled discourse dataset to create a model for industrial robots to engage in task-oriented dialogue with humans in real-world factory settings \cite{li2022tod}. 
While the authors conducted some preliminary user evaluations on their system's generated conversational responses, they only looked at some user perceptions to compare conversational responses generated by different model types. They did not study user behaviors in response to the system. 
In this work, we specifically attempt to understand how people will respond to a robot manipulator that engages in small talk with them during a collaborative task.

\subsection{People Treat Machines Like People}

The inclination of humans to treat machines socially can be contextualized within broader theoretical frameworks. The Computers Are Social Actors (CASA) paradigm \cite{nass1994computers} and the Media Equation Theory \cite{reeves1996media} argue that humans often, even subconsciously, treat machines, computers, and various forms of media as sentient entities, reacting to them in inherently social manners. 
Even in industrial manufacturing environments, an intrinsic human desire for social interaction remains. 
As operators often partake in casual conversations or ``small talk'' during shifts \cite{cheon2022robots}, 
they have also found themselves speaking to the robot they work with \cite{Sauppé_Mutlu_2015}. 
This suggests that individuals expect and actively desire social features in manufacturing robots.
Given the limited research on designing social behaviors for industrial robots and manipulators with a non-human form, despite its potential, a pressing question arises: 
How can we further design human-like attributes for robots that are inherently devoid of such characteristics? Building on this notion, our study investigates the potential of using small talk to instill social characteristics in a physical robot arm manipulator, specifically within the context of a human-robot collaborative assembly task.

\subsection{Small Talk}

Social dialogue that consists of light conversation between parties is often referred to as ``small talk'' \cite{Laver1981LinguisticRA}. This discourse may occur among individuals who have recently met or lack familiarity with each other, within settings such as a supermarket checkout, among colleagues who share some degree of acquaintance, or amidst family gatherings. \cite{coupland2014small}. Small talk is significant in the workplace as it is used by coworkers to establish trust, build rapport and strengthen interpersonal relationships \cite{coupland2003small, pullin2010small}. It typically occurs at the start of conversations with topics such as weather, sport, or other social engagements outside of the workplace \cite{holmes2005small}. 
While prevalent in American culture, attitudes toward the value of small talk differ across cultures. 
In Australia, some Chinese immigrant professionals were found to be unfamiliar with the dynamics and nature of small talk as it is unnecessary for them to regulate interpersonal relationships in Chinese society \cite{cui2015small}. Moreover, in Germany, a simple ''How are you?'' would be taken literally as it is not common practice in German culture to greet complete strangers in an attempt to initiate light small talk \cite{rings1994beyond,bickmore2005social}.

The integration of small talk to non-human agents engaging in conversations with people has been explored to help promote trust and rapport between the human and the agent. Small talk has mainly been applied to support additional conversations in settings where people are interacting with conversational agents \cite{clark2019good, feine2019taxonomy, kluwer2011-like}, voice assistants \cite{mahmood2023llm}, and virtual agents \cite{zhou2019trusting}. 
Additionally, incorporating small talk in more anthropomorphic social robots, combined with gaze \cite{babel2021small}, facial expressions \cite{paradeda2016facial}, or emotional displays \cite{aroyo2018trust}, has been shown to elevate user trust levels. 
However, the domain of applying small talk to foster rapport between humans and other agents with physical capabilities who are not typically poised to engage in conversations with people has been less explored. 
For instance, preliminary investigations conducted in a controlled testing environment have explored how passengers interact with a conversational agent in an autonomous self-driving vehicle, including observations of how the agent initiates greetings and engages in small talk conversations with riders \cite{large2019s}. 
Additionally, another group of researchers created a model based on an industrial-oriented Wizard-of-Oz conversation dataset for robots, which mixes task-oriented dialogue with small talk, that can generate system actions and conversational responses and predict belief states to facilitate human-robot collaboration in real-world industrial settings \cite{li2022tod}. 
While this preliminary study points to the feasibility of integrating small talk with task-oriented dialogue in manufacturing settings, its evaluations focused on 
some user perceptions of the generated conversational responses to compare models that utilized their new dataset rather than placing an emphasis on understanding how people behaved and responded to their overall system.
Further exploration is needed to understand how individuals interact with and utilize such robots to perform collaborative tasks.

\section{Methods}

We conducted an in-person study in which participants physically collaborated with an industrial robot equipped with small talk conversational capabilities. 
Intrigued by how individuals would interact with the non-human-like robot's small talk, we aimed to explore three general aspects of the overall interaction: people's reactions and responses to the robot's small talk, the general behaviors people exhibited during the small talk conversations, and participants' reactions to our introduced manipulation of negative team feedback.

\subsection{Study Task: Collaborative Assembly}

\begin{figure}[t]
  \includegraphics[width=\columnwidth]{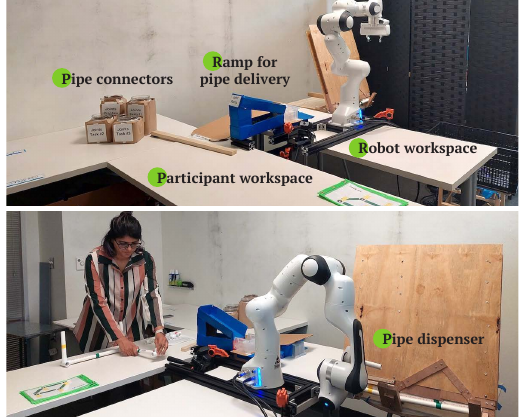}
  \caption{The experimental setup. The robot can grab a pipe from the pipe dispenser and place it on a pipe ramp, which delivers it to the participant. The participant can grab a joint from the joint containers on the left and place a faulty joint on the joint ramp for it to be retrievable by the robot.}
  \Description{The experimental setup. The top image shows two tables next to one another. The table on the left is the participant workspace and contains four boxes with pipe connectors that the participant can grab joints from. The table on the right is where the robot is mounted and serves as the robot's workspace. A ramp that the participant can use for the delivery of faulty pipes to the robot and a ramp that the robot can use to deliver requested pipes to the participant connect the two tables. The bottom image shows another view of the two tables, with a closer view of the robot workspace that highlights a pipe dispenser to the right of the robot from which the robot collects the pipe the participant requested.}
  \label{fig:workspaces}
\end{figure}

Participants were informed that the assembly task was intended to mimic a real-world collaboration setup for a human-robot team where the robot retrieves parts out of reach for the human and refrains from entering the human's designated workspace. They were given a fictional team assignment (Team 3) and were told the collaborative task scenario was a prototype for a larger-scale warehouse interaction. It was intended to simulate a modern warehouse that may contain a computerized efficiency monitoring system to oversee the team collaboration. The monitoring system would communicate with them via speech but would be distinguishable from any speech communication coming from the robot. The experimental setup is shown in Figure \ref{fig:workspaces}.

Each participant engaged in a collaborative PVC pipe structure building task with a robot manipulator mounted on a table. They were given diagrams corresponding to the structures they were required to assemble. The participants were asked to wear a microphone that would enable them to communicate with the robot. They were given instructions for how to interact with the robot via speech commands such as \emph{``Bring me a [green or yellow] pipe''}, in order for the robot to retrieve the corresponding pipe. The robot would provide a verbal confirmation message if a request was received correctly and an additional instruction reminder message once per structure when the participant requested their second to last pipe for that structure. The pipe would be placed on a ramp that connected the robot's workspace to the participant's workspace so that the pipe could be delivered to the participant without the robot entering their workspace. 
Participants were provided with joints in containers that corresponded to each structure to be built. To simulate a realistic industrial collaborative environment that might involve the use of defective materials, each participant was given one faulty joint for every pipe structure. They were instructed to inspect each joint before using it in the structure. If participants identified a faulty joint (marked with red sticker), they would place the joint on a ramp that would deliver it to a container inside the robot's workspace. The robot would retrieve the joint when given a verbal command such as \emph{``Report a faulty joint''}.

\subsection{System and Robot Behavior}
We present the system and robot behavior manipulations implemented in our study to understand how the social facets of a robot manipulator influence users' behaviors and perceptions of the robot.

\subsubsection{Negative Team Feedback}

Midway through the first and second structures in the main session, the human-robot team was provided with verbal negative feedback from the monitoring system; see Figure \ref{fig:task-windows}. 
Each message was prefaced by an audible beep: 
\begin{enumerate}
    \item[] Feedback 1: \emph{``System announcement. Team 3, please work on improving your performance.''}
    \item[] Feedback 2: \emph{``System announcement. Team 3, you are currently not meeting expectations. Improve your performance.''}
\end{enumerate}

\subsubsection{Robot Social Behavior: Small-Talk}

The robot exhibited two types of social speech behaviors: questions and statements, and two types of task-oriented speech behaviors: reminders and confirmations.
During the main assembly task, the robot would engage in small talk by initiating questions and/or statements after it delivers a request confirmation message or an instruction reminder message.
The robot initiated a minimum of one question or statement per pipe request, regardless of the response rate of the participants. After a break in conversation via the negative feedback announcement, the robot initiated questions and statements on a different topic. 
The conversation touched on a range of questions spanning three primary topics to accommodate the diverse interests of individuals: 1) travel; 2) books, TV shows, and movies; and 3) sports and games. Table \ref{tab:stalk} shows some examples of the robot's small-talk phrases.

\begin{table}[t]
    \centering
    \caption{Examples of robot phrases that are questions (non-shaded) and statements (shaded grey). }

    \label{tab:stalk}
    \begin{tabular}{l|l}
    \hline
    \textbf{Topic}                       & \textbf{Phrase}                                                                     \\ \hline
                                         & \textit{By the way, do you like to travel?}                                         \\ \cline{2-2} 
    \multirow{-2}{*}{Travel}             & \cellcolor[HTML]{EFEFEF}\textit{I arrived to the U.S. from Germany in a large box.} \\ \hline
                                         & \textit{What do you like about those books?}                                        \\ \cline{2-2} 
    \multirow{-2}{*}{Books, TV,  Movies} & \cellcolor[HTML]{EFEFEF}\textit{That sounds like a great way to relax.}             \\ \hline
                                         & \textit{Do you have a favorite sport to watch?}                                     \\ \cline{2-2} 
    \multirow{-2}{*}{Sports, Games} & \cellcolor[HTML]{EFEFEF}\textit{It is good to spend time outdoors exercising.} \\ \hline
    \end{tabular}
\end{table}

\subsection{System Implementation}

\subsubsection{Robot System}
We utilized a Panda robot from Franka Emika. 
It is an industrial robotic arm with 7 degrees of freedom (7-DoF). 
The robot was restricted to its designated workspace and never intruded into the participant's allocated area. 
Its movements were preprogrammed to execute specific sequences of actions. 
Among these, two involved pick-and-place motions: one to retrieve and deliver either a green or yellow PVC pipe to the participant, and the other to pick up a container holding a PVC pipe joint and subsequently move the container over a bin to discard the joint. 
The robot interacted with participants using text-to-speech communication.
Its speech and physical actions were controlled by the experimenter via a Wizard-of-Oz (WoZ) approach \cite{woz2012riek}.

\subsubsection{Wizard-of-Oz Control} 
While monitoring the study interaction using a webcam, the experimenter used a remote graphical user interface (GUI) to click buttons that executed preprogrammed robot actions. 
For conversational speech, pre-written conversational phrases were assigned to specific alphanumeric codes. The experimenter could trigger one of the preprogrammed conversational speech phrases by inputting its code into a textbox in the GUI. 
Additionally, the experimenter could manually type text into the textbox to produce non-preprogrammed phrases via Google Text-to-Speech (TTS). 

\begin{figure}[t]
  \includegraphics[width=\textwidth]{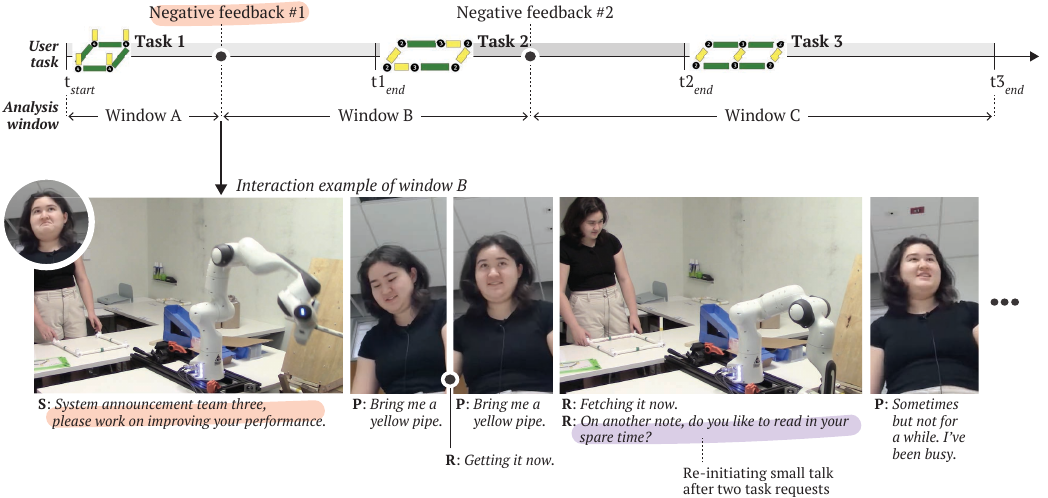}
  \caption{An overview of the collaborative tasks the participants completed with the robot. The team negative feedback was introduced halfway through the first and second task structures. To analyze changes in user behavior across instances of negative feedback, we delimit observation windows with respect to the negative feedback announcements. Window A is considered our baseline as this corresponds to the only portion of the interaction that has not yet been influenced by any negative feedback statements. In the lower half of the figure, we include an example of one user's conversational interaction with the robot after the first negative feedback announcement (P10).}
  \Description{Top half of the figure contains an outline of the three tasks, with dots on a timeline signifying where a negative feedback announcement was given. The bottom of the figure contains images of the participant alongside their and the robot's conversation as the task progresses. Next is the caption of the conversation, with S corresponding to the system announcement, R corresponding to the robot, and P corresponding to the participant; S: System announcement, team three, please work on improving you performance. P: Bring me a yellow pipe. R: Getting it now. P: Bring me a yellow pipe. R: Fetching it now. R: On another note, do you like to read in your spare time? (re-initiating small talk after two task requests) P: Sometimes but not for a while. I've been busy.}
  \label{fig:task-windows}
\end{figure}

\subsubsection{Preprogrammed Speech Behavior}
When a robot movement was triggered from the GUI, a confirmation message was automatically activated to notify the participant that the action would occur. 
Robot confirmation messages were preprogrammed and selected randomly during each execution. 
Task reminder messages were similarly preprogrammed and triggered automatically when the participant had requested enough pipes to be one away from completing a structure. 
Negative feedback directed towards the human-robot team was triggered automatically when participants had requested half the pipes needed for the first and second structure; see Figure \ref{fig:task-windows}.

\subsection{Study Procedure}

Participants first completed the consent form and a demographics questionnaire. 
Afterward, the experimenter explained the task scenario and gave the participant instructions about the task and how to communicate with the robot. 
After the experimenter left the room, the participant received an introductory message from the monitoring system. 
The participant then engaged in a practice session to construct one pipe structure, allowing them to become acquainted with the robot's task-oriented behaviors and the use of verbal commands. 
The robot only exhibited task-oriented dialogue during the practice task, such as a reminder message to move their pipe structure over to another table when finished and a confirmation message signifying it comprehended the request. 
Following the practice session, they took part in a main session where they built three additional pipe structures with the robot. 
At the end of the study, 
the experimenter conducted a semi-structured interview with each participant to better understand their collaborative experience. 

\subsection{Data Analysis} 

We collected audio-visual data of the participants' collaboration with the robot during the assembly tasks, and of the semi-structured interview.
The experimenter transcribed the audio data with the help of audio transcription software. 
Each participant had an average of 12 minutes of interaction with the robot over the main tasks.
The experimental interaction was divided into three windows (see Figure \ref{fig:task-windows}) with respect to the system's negative feedback messages:

\begin{itemize}
    \item Window A. From Start to Feedback Announcement 1
    \item Window B. From Feedback Announcement 1 to Feedback  Announcement 2
    \item Window C. From Feedback Announcement 2 to End
\end{itemize}

To the best of our knowledge, we found no existing coding schemas applicable to an industrial robot that exhibits small talk, resulting in the development of our own codebook to identify central themes and patterns surrounding the participants' interactions with the robot. 
We performed thematic analysis on the transcripts using a combination of inductive and deductive coding. 
Inductive coding was used to create the initial codebook. 
When the codebook was finalized, a deductive approach was used to code the transcriptions. 
The phrases spoken by the robot and participant during the main session were coded by the context in which it was made, such as a task-oriented message or a small talk response. 
The codebook can be viewed in Table \ref{tab:code_overview_a} and Table \ref{tab:code_overview_b} in the Appendix.

Two coders were used to code the study data. 
The primary coder first coded the entire data set, and the secondary coder then coded 10\% for reliability.  
After resolving disagreements between two coders via discussion, we found almost perfect inter-coder agreement on 10\% of our categorical and continuous data, as evidenced by a Cohen’s Kappa of 0.98 and an Intraclass Correlation Coefficient (ICC3) of 0.98.

\subsection{Measures} 

\begin{table}[t]
\caption{Examples of People's Responses That Vary in Length}
\label{tab:length}
\begin{tabular}{|l|l|}
\hline
\textbf{Category}                & \textbf{Conversation}                                    \\ \hline
 &
  \textit{\begin{tabular}[c]{@{}l@{}}R:  Do you like to play sports?\end{tabular}} \\ \cline{2-2} 
\multirow{-2}{*}{Short Sentence}        & \cellcolor[HTML]{FFF0B5}\textit{P: No, not really. (P6)}         \\ \hline
                                 & \textit{R: How often do you watch movies?}               \\ \cline{2-2} 
\multirow{-2}{*}{Medium Sentence} &
  \cellcolor[HTML]{FFF0B5}\textit{\begin{tabular}[c]{@{}l@{}}P: Not that often, like twice three times \\ a month. (P20)\end{tabular}} \\ \hline
                                 & \textit{R: Where have you been in the United States?}    \\ \cline{2-2} 
\multirow{-2}{*}{Multi-Sentence} &
  \cellcolor[HTML]{FFF0B5}\textit{\begin{tabular}[c]{@{}l@{}}P: Most of the East Coast, a lot of the West\\ Coast, the Midwest. I was born in Utah. Uh, I've\\ been to a lot of places in the United States. (P6)\end{tabular}} \\ \hline
\end{tabular}%
\end{table}

We developed a set of metrics to break down participants' involvement in small talk with the robot throughout the collaborative interaction, as well as their responses and reactions to both the small talk and the negative team feedback. 

\subsubsection{Initial Reactions to Robot Small Talk}
Given the robot's non-human-like appearance, people might find its small talk unexpected.
Thus, we decided to analyze the users' reaction to the first robot small talk conversational attempt to help us understand the impact of small talk in robots primarily designed for physical collaboration. 
\begin{itemize}
    \item \textbf{Conversation reciprocation.} We coded a person's response to the robot's first conversational attempt as no response or response, and checked if the user's response was an answer to a robot question.
    \item \textbf{Task disruption.} We classified people's behavioral response to the robot's first small talk question as one of three actions: minimal (quick glance at the robot but no task disruption), freeze (stops current action and looks at the robot before continuing the task), and none (did not stop or look at the robot). 
    \item \textbf{Affective reactions.} To gather insights into the participants' emotional or mental state, we grouped people's non-verbal affective reactions during the robot's first question into three categories: pleasant (smile), surprise (chin raise, lower lip raise, brow lowering etc.), and indifference (no reaction).
\end{itemize}


\subsubsection{User Engagement with Robot Small Talk}


The participants' small talk responses were coded to identify the general length of the response; see Table \ref{tab:length} for examples. Individuals' responses that consisted of a participant-initiated question to the robot were separately identified and labeled accordingly; see Table \ref{tab:code_overview_a}. 

\begin{itemize}
     \item \textbf{Length of participant response.} A user response that was a \textit{short sentence} consisted of $<= 5$ words, a \textit{medium sentence} consisted of $<= 5$ words, and a\textit{multi-sentence} response consisted of more than one sentence.
\end{itemize}
Furthermore, we observed whether the participants' verbal responses were a reply to a robot small talk question, small talk statement, reminder message, or confirmation message. We thus evaluate the users' response rates to the robot's verbal phrases based on context using the following:

\begin{itemize}
    \item \textbf{Response Rate.} The percentage of robot
    messages that elicited a response from the user. This metric is broken down into \textit{Task-Oriented Response Rate}, which pertains to the robot task-oriented confirmation and reminder messages, and \textit{Small Talk Response Rate}, which pertains to robot small talk questions and statements. 
\end{itemize}
We were also curious as to whether people would be willing to extend the small talk conversation with the robot if given the opportunity to do so. We thus evaluate the frequency at which individuals extended their conversation with the robot using the following metric: 
\begin{itemize} 
    \item \textbf{Conversation Extension Rate.}  A percentage calculated by dividing the total number of user responses directly prompted by a robot small talk statement and the total number of user-initiated questions by the total number of robot small talk phrases.
\end{itemize}

\subsubsection{Participant Reactions to System Negative Feedback}
We were intrigued by how individuals may perceive external negative feedback from within their human-robot team collaboration, especially as the robot engages in small talk. Our general interest lied in understanding participants' reactions to the feedback, how they attribute blame, and whether their behavior towards the robot changes post-feedback delivery. Thus, we examine the following aspects: 

\begin{itemize}
    \item \textbf{Verbal or non-verbal response.} We analyzed a time window starting from the system feedback announcement and lasting up to 5 seconds post-announcement, yielding a duration of 10 seconds for the first and 13 seconds for the second announcement. For each window, we observed the participants' reactions in response to the negative feedback announcements. 
    \item \textbf{Blame attribution.} Participants were asked in the post-study interview: \textit{What do you think was the cause of the system announcement?} This was followed by a question to categorically measure blame attribution for the negative feedback: \textit{Who would you say is at fault for the system announcement? No one, the team, you, the robot?}
    \item \textbf{Participant Response Patterns.} We categorized participants' small talk responses into one of three different participant response patterns to observe how participants naturally alternate between small talk and task requests: \textit{small talk before task request}, \textit{small talk after task request}, and \textit{halted small talk}. A halted small talk response pattern occurred when a user began their small talk response, paused to make a task request, and then resumed their response after. 

\end{itemize}

\subsection{Participants}
There were 24 participants recruited from the authors' university campus, out of which 4 were excluded due to technical issues (\eg computer system froze, speakers malfunctioned, text-to-speech API failed). 
The remaining 20 participants (11 female, 9 male) ranged in age from 18 to 38 ($M = 24.35, SD = 5.44$). Twelve participants listed Asian as their ethnicity; five as Caucasian; one as Hispanic, Latino, or Spanish origin of any race; one as Asian and Caucasian; and one as Caucasian and Hispanic, Latino, or Spanish origin of any race. 
Seventeen participants selected the United States as their current country of residence; one selected Taiwan; one selected Hong Kong; and one selected Germany. 
While all participants spoke English, three listed English as their only spoken language, with seventeen indicating they speak two or more languages.
When asked about their native language, ten participants listed English; five listed Mandarin; one listed Korean; one listed Hindi; one listed Tamil; one listed German, and one listed Spanish.
Participants had a variety of educational backgrounds, including computer science, engineering, natural sciences, education, biotechnology, administration, and international studies. Participants were familiar with technology ($M = 4.15, SD = 0.75$, 5-point scale with 5 being Very Familiar and 1 being Not At All), had moderate experience with robots ($M = 2.5, SD = 1$), and had little experience teaming with robots ($M = 1.65, SD = 0.93$). 
The study took about 45 minutes and was approved by the university's institutional review board. Participants were compensated with \$15 US dollars. 

\section{Results}
\begin{figure}[t]
  \includegraphics[width=0.75\columnwidth]{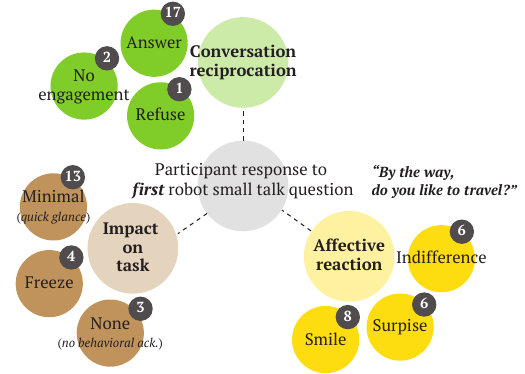}
  \caption{Different aspects of participants' response to the first robot small talk initiation:  (1) conversational reciprocation, (2) task disruption, and (3) affective reaction. For each aspect, the states are mutually exclusive. The numbers denote participants who exhibited the corresponding responses.}
  \Description{Figure 3: A figure graphically displaying the participant's response to the first robot small task question (e.g., "By the way, do you like to travel?") in terms of conversation reciprocation, impact on task, and affective reaction, with numbers listed for each type of participant response (conversation reciprocation: 17 answer, 2 no engagement, 1 refuse; task disruption: 13 minimal [quick glance], 4 freeze, 3 none [no behavioral acknowledgment]; affective reaction: 6 indifference, 6 surprise, 6 smile)}
  \label{fig:results-first-response}
\end{figure}

This work strives to understand whether, how, and to what extent participants engage with robot-initiated small talk. 
We first assessed how participants perceived the robot's initial small talk question, specifically its impact on their ongoing actions and emotional state. 
Next, we examined people's engagement in small talk by looking at their types of responses, the rate at which they responded to the robot's small talk or task-oriented phrases, and if and when they decided to extend the small talk conversation. 
Third, we observed how individuals reacted to the system's negative feedback announcements towards the human-robot team, and how it affected their perception of the team's collaboration. 

\subsection{Initial Reactions to Robot Small Talk}
\label{results:first-question}
We first examine participant reactions to the robot's first small talk question and present our findings on three different aspects of their responses; see Figure \ref{fig:results-first-response}. 
\subsubsection{Conversation Reciprocation}
Eighteen participants responded to the robot's first question: \textit{``By the way, do you like to travel?''}; only two participants did not respond to the robot's question; the same two participants did not engage in any non-task conversation with the robot. 
Of the eighteen respondents, seventeen responded with a non-question response: thirteen with a short sentence and four with a medium sentence, each containing an answer similar to, for example, \textit{``Yeah, I like traveling.''} 
One of the eighteen participants, however, replied with a rhetorical question: \textit{``Why are you talking to me robot? This has nothing to do with the task.''} (P12). 

\subsubsection{Task Disruption}
The majority of participants ($n=13$) merely exchanged a brief glance with the robot when it posed the first question without letting it interrupt their ongoing task, suggesting that their attention was momentarily captured by the robot's ``unusual'' behavior (initiating small talk), but not to a degree that proved disruptive. In contrast, three participants did not look at the robot at all. Interestingly, four participants momentarily halted their task for at least two seconds, looked intently at the robot (indicating a \textit{freeze} reaction), and then resumed their activity; these individuals also engaged in a verbal response to the robot. 

\subsubsection{Affective Reaction}
Eight out of the 20 participants smiled in response to the robot's first question, indicating a pleasant emotional state. Six participants displayed signs of surprise, as indicated by two coders, through various non-verbal cues such as chin raiser, lower lip raiser, and eye-brow lowering. The remaining six participants exhibited indifference, as evidenced by their lack of any non-verbal response.

\subsection{User Engagement with Robot Small Talk}

\subsubsection{Length of Participant Responses}
\label{results:length}

A total of 377 non-question verbal responses were made to the robot for non-task-oriented dialogue; examples are shown in Table \ref{tab:length}. Participants' responses were mostly one sentence ($n=319$, $84.62\%$). The one-sentence responses were further broken down as follows: short sentence (<= 5 words; $n=179$, $56.11\%$), and medium-length sentence (> 5 words; $n=140$, $43.89\%$). The remaining participant responses comprised multiple sentences ($n=58$, $15.38\%$), reflecting higher and possibly meaningful engagement.


\begin{table}[t]
\caption{Response rates of participant responses made directly to the robot's speech by robot speech type.}
\label{tab:response-rates}
\resizebox{\textwidth}{!}{%
\begin{tabular}{rrcccccc}
\multicolumn{1}{l}{} &
  \multicolumn{1}{l}{} &
  \multicolumn{1}{l}{} &
  \multicolumn{1}{l}{} &
  \multicolumn{1}{l}{} &
  \multicolumn{1}{l}{} &
  \multicolumn{1}{l}{} &
  \multicolumn{1}{l}{} \\ \cline{3-8} 
\multicolumn{1}{l}{} &
  \multicolumn{1}{l|}{} &
  \multicolumn{3}{c|}{\cellcolor[HTML]{84BE42}\textbf{Robot Small Talk}} &
  \multicolumn{3}{c|}{\cellcolor[HTML]{E2D4C4}\textbf{Robot Task-Oriented}} \\ \cline{3-8} 
\multicolumn{1}{l}{} &
  \multicolumn{1}{l|}{} &
  \multicolumn{1}{c|}{\cellcolor[HTML]{84BE42}\textit{Questions}} &
  \multicolumn{1}{c|}{\cellcolor[HTML]{84BE42}\textit{Statements}} &
  \multicolumn{1}{c|}{\cellcolor[HTML]{84BE42}\textit{Combined}} &
  \multicolumn{1}{c|}{\cellcolor[HTML]{E2D4C4}\textit{Confirmations}} &
  \multicolumn{1}{c|}{\cellcolor[HTML]{E2D4C4}\textit{Reminders}} &
  \multicolumn{1}{c|}{\cellcolor[HTML]{E2D4C4}\textit{Combined}} \\ \hline
\multicolumn{1}{|r|}{} &
  \multicolumn{1}{r|}{\textit{Percentage}} &
  \multicolumn{1}{c|}{83.87\%} &
  \multicolumn{1}{c|}{44.62\%} &
  \multicolumn{1}{c|}{70.79\%} &
  \multicolumn{1}{c|}{1.48\%} &
  \multicolumn{1}{c|}{18.33\%} &
  \multicolumn{1}{c|}{3.37\%} \\ \cline{2-8} 
\multicolumn{1}{|r|}{\multirow{-2}{*}{\textbf{\begin{tabular}[c]{@{}r@{}}All Participants \\ (n = 20)\end{tabular}}}} &
  \multicolumn{1}{r|}{\cellcolor[HTML]{EFEFEF}\textit{Ratio}} &
  \multicolumn{1}{c|}{\cellcolor[HTML]{EFEFEF}(312 / 372)} &
  \multicolumn{1}{c|}{\cellcolor[HTML]{EFEFEF}(83 / 186)} &
  \multicolumn{1}{c|}{\cellcolor[HTML]{EFEFEF}(395 / 558)} &
  \multicolumn{1}{c|}{\cellcolor[HTML]{EFEFEF}(7 / 474)} &
  \multicolumn{1}{c|}{\cellcolor[HTML]{EFEFEF}(11 / 60)} &
  \multicolumn{1}{c|}{\cellcolor[HTML]{EFEFEF}(18 / 534)} \\ \hline
\multicolumn{1}{|r|}{} &
  \multicolumn{1}{r|}{\textit{Percentage}} &
  \multicolumn{1}{c|}{88.64\%} &
  \multicolumn{1}{c|}{47.98\%} &
  \multicolumn{1}{c|}{75.24\%} &
  \multicolumn{1}{c|}{1.64\%} &
  \multicolumn{1}{c|}{20.37\%} &
  \multicolumn{1}{c|}{3.75\%} \\ \cline{2-8} 
\multicolumn{1}{|r|}{\multirow{-2}{*}{\textbf{\begin{tabular}[c]{@{}r@{}}ST Participants \\  (n = 18)\end{tabular}}}} &
  \multicolumn{1}{r|}{\cellcolor[HTML]{EFEFEF}\textit{Ratio}} &
  \multicolumn{1}{c|}{\cellcolor[HTML]{EFEFEF}(312 / 352)} &
  \multicolumn{1}{c|}{\cellcolor[HTML]{EFEFEF}(83 / 173)} &
  \multicolumn{1}{c|}{\cellcolor[HTML]{EFEFEF}(395 / 525)} &
  \multicolumn{1}{c|}{\cellcolor[HTML]{EFEFEF}(7 / 426)} &
  \multicolumn{1}{c|}{\cellcolor[HTML]{EFEFEF}(11 / 54)} &
  \multicolumn{1}{c|}{\cellcolor[HTML]{EFEFEF}(18 / 480)} \\ \hline
\end{tabular}%
}
\end{table}

\subsubsection{Participant Response Rates}
\label{results:response-rates}
Of the 20 participants, two participants did not engage in additional conversational interaction with the robot apart from the verbal commands needed to complete the collaborative task. 
We determined the overall small talk response rate across all participants to be $77.60\%$ (433 out of 558); this total included any small talk questions or statements made by the participant.
Of the 18 participants who engaged in small talk, 17 responded to the robot's small talk throughout the experiment. One participant refrained from small talk after responding to the robot's second small talk question. 
We calculated the overall small talk response rate for the 18 participants who engaged in small talk to be $82.48\%$ (433 out of 525). 
We further calculated the participants' response rate towards the robot's small talk and non-task-oriented speech; this pertains to the robot phrases that directly elicited a user response. 
These results are summarized in Table \ref{tab:response-rates}. 


\subsubsection{Participants' Extension of the Conversation}
\label{results:user-initiated} 

\begin{figure}
    \includegraphics[width=\linewidth]{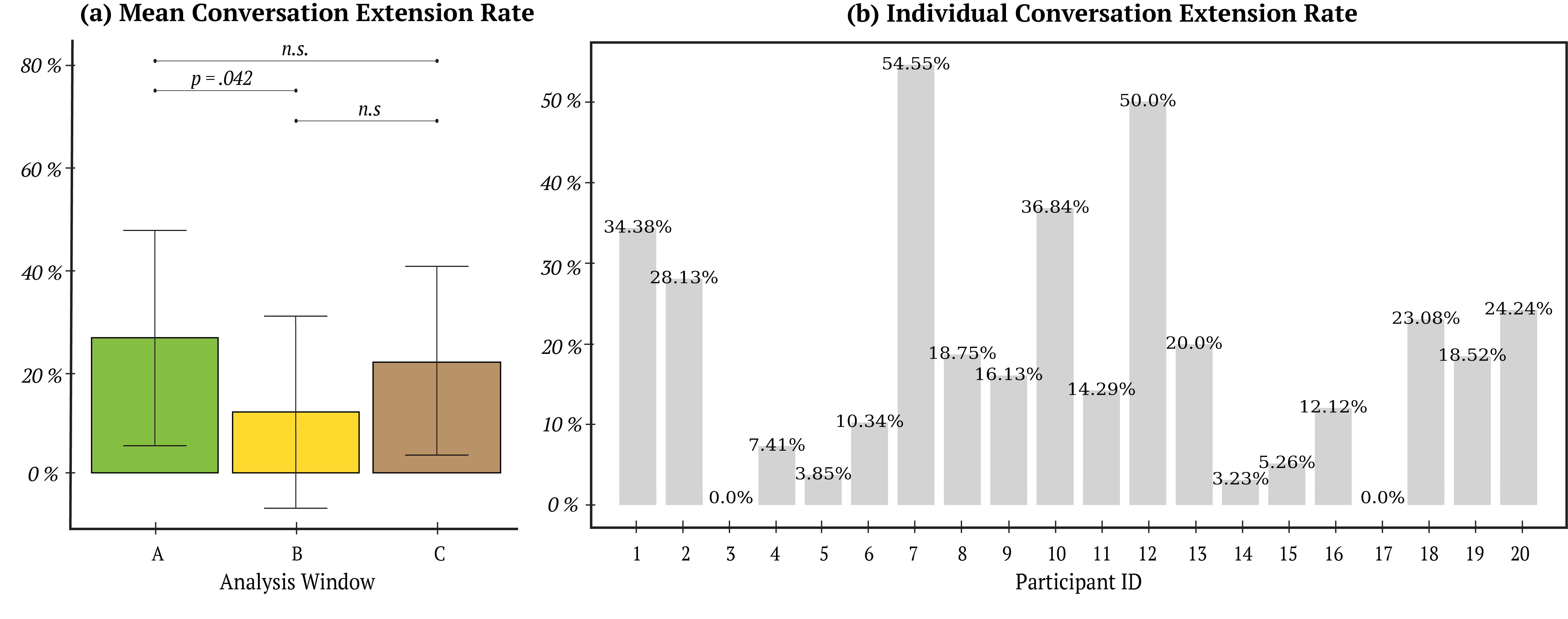}
    \caption{The conversation extension rate is a ratio of the number of participant responses and participant-initiated questions made in direct response to a robot small talk statement, over the total number of robot questions and small talk statements. (a) The average conversation extension rate across all participants over three separate time windows: Window A) before first negative feedback, Window B) between first and second negative feedback, Window C) after second negative feedback. A significant difference was found between the mean conversation extension rates of Windows A and B. (b) Scatter plot depicting the conversation extension rate for each individual participant across their entire interaction.}
    \Description{KP TO DO}
    \label{fig:extension}
\end{figure}

Of the 18 participants who engaged in small talk, all participants extended the conversation at least once, meaning they each either responded to one of the robot's small talk statements, or asked the robot a question. 
Each participant's average conversation extension rate is shown in Figure \ref{fig:extension}(b).

We noticed that more than half of the participants interacted with the robot by asking questions rather than just responding to the small talk initiated by the robot. Specifically, thirteen participants ($65\%$) asked a total of 56 questions. Out of these questions, some ($n = 14, 25\%$) were seeking clarification, with users asking the robot to repeat its previous question or statement. Only four user-initiated questions were direct responses to the robot's questions, without requesting a repetition of what was previously said.

The non-clarification questions ($n = 42, 75\%$) asked across nine participants were more personal, ranging from courteous ask-backs to inquiries about the robot's preferences or capabilities.
For example, P1 asked the robot, \textit{``Is there anywhere you want to travel?''}, later explaining in the post-study interview, \textit{``I wanted to know more about the robot, like once it started talking to me''}. Other participants also expressed genuine curiosity as a motivator. P7 remarked, \textit{``Because he's a robot and I'm not familiar with a robot and \dots I \dots truly want to know \dots how long does it take to get assemble \dots he said, \dots he's from Germany by a group, but I'm more \dots interested \dots I tried to ask \dots do you like your owner?''} Examples of other questions asked by P7 include, \textit{``And panda robot, do you do any sport?''}, which was later followed by, \textit{``So if you don't do sport, what do you do for your free time?''}
 
For the seven participants who did not initiate a question to the robot, many described their perception of the robot as non-human-like as a contributing factor. One participant commented how the conversation, \textit{``just doesn't have as much value as \dots if I was talking to an actual person and hearing where they want to travel''} (P13). For the two participants that did not engage with the robot at all, one stated \textit{``It's not a human, you don't have a conversation''} (P17). The other explained how \textit{``from beginning to end, I always felt like it's a robot''} (P3).

\subsection{Participant Reactions to System Negative Feedback}
\label{results:negative-feedback}

\subsubsection{Verbal and non-verbal responses.}

Responses to negative feedback statements were observed in all but two participants ($n=18$), who notably remained disengaged with the robot throughout the interaction.
Additionally, among the 18 participants who did respond to the feedback, two individuals reacted to only one of the two feedback statements.
Out of a total of 40 negative feedback messages, eight prompted verbal responses, while 34 evoked non-verbal reactions.
Of the eight verbal reactions, three were made towards the first negative feedback statement, and five towards the second.
However, these responses were spread across six unique participants, with two participants verbally reacting to both feedback messages. 
Among the four participants who responded verbally to only one of the two feedback messages, one reacted to the first message, and three to the second.    

Participants exhibited a range of non-verbal reactions; while some smiled and laughed, others gestured in frustration or hesitated, seemingly confused. Eye-rolling was also observed in response to the system announcement. Verbal reactions further highlighted this confusion. One participant remarked, \textit{``I ... okay''} (P18), while others expressed frustration and justified their own performance, such as \textit{``I'm doing as much as I can do, system announcer''} (P12) or the team's performance; \textit{``I mean we're fast enough''} (P7). These reactions are summarized in Table \ref{tab:react}.

\begin{table}[t]
\caption{Reactions in response to team negative feedback across both negative feedback announcements}
\label{tab:react}
\resizebox{0.6\columnwidth}{!}{%
\begin{tabular}{|clc|}
\hline
\textbf{Reaction}      &                                       & \multicolumn{1}{l|}{\textbf{Count}} \\ \hline
\multicolumn{1}{|c|}{\multirow{6}{*}{\textbf{Non-Verbal}}} & Brow movement (raise)                              & \textbf{3} \\ \cline{2-3} 
\multicolumn{1}{|c|}{} & Eye movement (\eg widen, blink)      & \textbf{10}                         \\ \cline{2-3} 
\multicolumn{1}{|c|}{} & Hand movement (\eg shake, scratch)   & \textbf{24}                         \\ \cline{2-3} 
\multicolumn{1}{|c|}{} & Head movement (\eg shake, nod)       & \textbf{19}                         \\ \cline{2-3} 
\multicolumn{1}{|c|}{} & Mouth movement (\eg smile, lip curl) & \textbf{25}                         \\ \cline{2-3} 
\multicolumn{1}{|c|}{} & Stands Still ( 2+ seconds)            & \textbf{14}                         \\ \hline
\multicolumn{1}{|c|}{\textbf{Verbal}}                      & \eg ``okay'', ``oh no'', ``I mean we're fast enough'' & \textbf{8} \\ \hline
\textbf{Total}         &                                       & \textbf{103}                        \\ \hline
\textbf{No reaction}   &                                       & \textbf{6}                          \\ \hline
\end{tabular}%
}
\end{table}

\subsubsection{Blame Attribution}
\label{results:blame}
Overall, 10 participants blamed no one for receiving the negative system feedback, 4 blamed the robot, 3 the team, and 3 themselves. 
Of those blaming the robot, only one cited its small talk while the rest attributed the announcements to its slow movement speed.
One participant who blamed the team stated, \textit{``\dots Maybe in terms of like the team chemistry, that was probably where it was lacking}'' (P16). Another explained how they thought the novelty of the conversation impacted the team's performance, \textit{``I also get a little more distracted, because \dots this is just \dots a new thing \dots it's something that I guess you would get used to if you're working with it every day}''(P10).
Those who blamed themselves acknowledged possible mistakes. Meanwhile, 15 participants discussed they were confused about what prompted the system announcements because they thought the team was performing efficiently. P1 explained how the announcement \textit{``felt like out of my control. So I was like, I don't think there was anything else that could have been done.''}

\subsubsection{Participant Responses to Robot Small Talk After Feedback}

Additionally, we were curious as to whether a negative evaluation of the human-robot team from the system influences the way participants respond to the robot's small talk; we investigated this by utilizing the defined interaction windows (see Figure \ref{fig:task-windows}) to compare differences in response rates and conversation extension rates across the windows for each participant.
We conducted a Kruskal-Wallis Test to explore whether the negative system feedback affected participants’ response rates to robot questions and statements but found no significant differences across the three interaction windows $({\chi}^2 (2) = 0.76$, $p$ = $.683)$. 
The average conversation extension rate per participant for the participants who engaged in small talk ($n=18$) was $(M = 21.17\%, Mdn = 18.63\%, SD = 14.91\%)$.
We found a statistically significant difference in the  average conversation extension rate over the three windows as revealed by a Kruskal-Wallis Test (${\chi}^2 (2) = 6.91, p = .032$)
A post hoc Dunn's test of multiple comparisons showed a statistically significant decrease $(p = .042)$ in a participant's average conversation extension rate between Windows A $(M = 26.8\%, SD = 21.1\%)$ and B $(M = 12.2\%, SD = 18.8\%)$, which are before and after the first feedback announcement. See Figure \ref{fig:extension}(a). 
We considered $p$ < 0.05 a significant effect, and selected the Kruskal-Wallis Test due to assumptions of a One-Way ANOVA not being met and subsequently used the Dunn Test, if applicable, to control for multiple hypothesis testing.

\begin{figure*}
  \includegraphics[width=\textwidth]{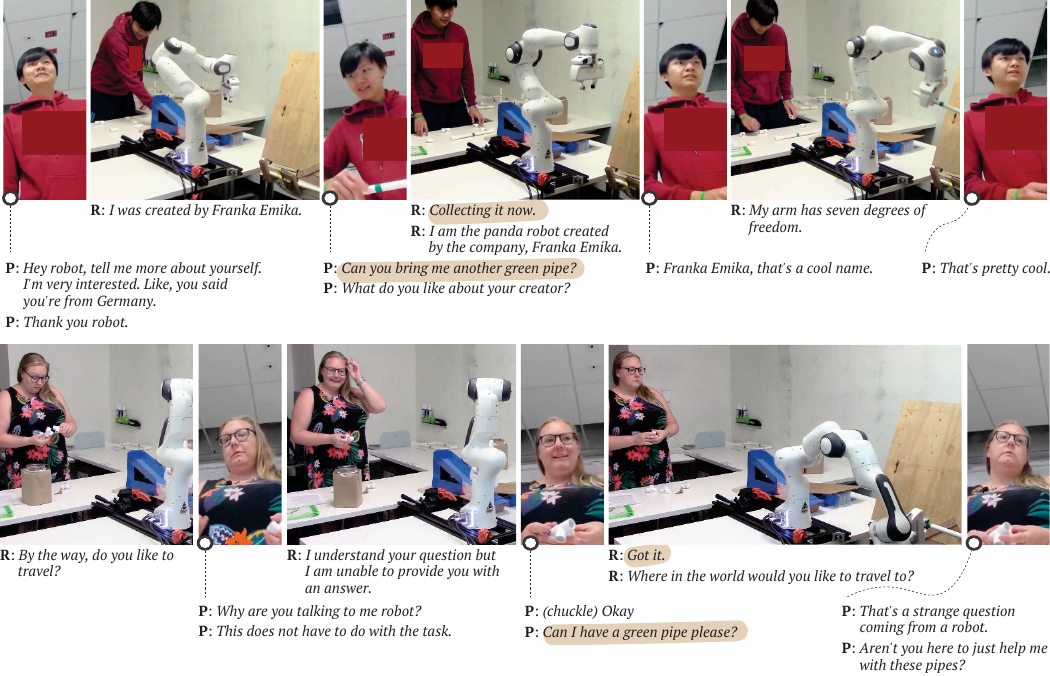}
  \caption{Examples of positive and negative user interaction with the robot small talk behaviors. The top row shows a positive example where the participant expressed genuine interest in learning more about the robot (P7). The bottom row is an example of a user who perceived the interaction negatively and exhibited a potential mental model mismatch of expectations (P12). }
  \Description{Figure 4: Examples of positive and negative user interaction with the robot small talk behaviors. The figure has two rows. The top row contains seven images alternating between images focusing on the participant and images focusing on the robot, with snippets of the conversation shown below indicating participant interest in conversation about the robot's background and specifications. The bottom row contains six images alternating between images focusing on the participant and images focusing on the robot, with snippets of the conversation shown below indicating participant indicating surprise or confusion on why the robot is engaging in conversation unrelated to the task at hand.}
  \label{fig:pos-neg-example}
\end{figure*}

\section{Discussion}








We explored the incorporation of conversational elements such as small talk into industrial robots, especially those not inherently built for social interactions. 
While opinions on the robot's conversational abilities vary, participants showed a willingness to engage in off-topic conversations initiated by the robot. 
In this section, we discuss our findings in relation to prior work, identify design considerations for future conversational interactions, and acknowledge the limitations of our current work alongside variables that warrant further exploration. 

\subsection{Social Conversations in Physical Human-Robot Collaboration}

\label{discussion:meaningful}


Prior research indicates that people in manufacturing settings value social interactions; specifically, robot operators engaged in small talk at times with fellow operators during their shifts \cite{cheon2022robots}.
As it is common for humans to partake in human-human interaction with their co-workers, it is an interesting observation that humans may find themselves talking to their robot partners. 
Additionally, a different set of robot operators revealed in interviews that they occasionally spoke to the robot and expressed a desire for the robot to exhibit more social behaviors \cite{Sauppé_Mutlu_2015}. 

When asking whether or not we should include conversational robot interactions in manufacturing settings, we have to recognize that people are inclined to speak with one another and with their robot counterparts.  
This tendency to speak to the robot aligns with the evidence discovered in this work (see Sec. \ref{results:response-rates}). Most of our participants engaged in casual conversations with their robot assistant; they responded to robot small talk arising from a traditionally non-social robot during a non-social collaborative assembly task. 

In our study, the participants who exhibited non-task-oriented conversations while interacting with the robot shared different incentives for engaging with it.  
One individual explained their engagement may have been due to reciprocity, \textit{``It initiated questions to me. So I was like, Oh, I guess I can talk to it... I mean ... if something's talking to me, I feel like I typically talk back''} (P10). 
However, many participants presented personal questions to the robot due to their curiosity about its functionalities and background (see section \ref{results:user-initiated}). 
For instance, as illustrated in Figure \ref{fig:pos-neg-example} top, the participant inquired, \textit{``Hey robot, tell me more about yourself. I'm very interested. Like, you said you're from Germany.''} 
Another participant, referencing Figure \ref{fig:pos-neg-example} bottom, sought clarity on the robot's primary function by asking, \textit{``Aren't you here just to help me with these pipes?''}.
It seems as if the robot's display of its verbal capabilities was an invitation for the user to communicate in return with a verbal response.

The participants' willingness to advance the conversation beyond simple or short responses is demonstrated by their medium and multi-sentence responses, resulting in about 44\% and 15\% of their social responses that were not a question (see section \ref{results:length}). 
Users displayed a noticeable disposition to respond to the robot's direct questions, provide supplemental information, and reply to the robot's comments that did not warrant a user response. 
This level of user engagement is evident in Figure \ref{fig:teaser}, where a participant's response goes beyond merely answering the robot's query, not only describing aspects of the discussed game (Tetris) but anticipating the robot's genuine interest in the game by commenting (\textit{``You might like it.''}) on the robot's response (\textit{``No, I have not''}) to her question (\textit{``if you've ever played that?''}). 
When describing the game's premise in additional detail, the participant not only made eye contact with the robots' body but also used hand and arm gestures to supplement their explanation. 
We found this participant's use of gestures a bit surprising, as the robot does not have a face, we did not place cameras on the robot itself, and we did not inform participants with details about the robot's visual processing capabilities. 

Overall, it appears that participants wanted to talk to the robot and provide more meaningful responses during their conversations.
We thus are making a case for more social communication with robots during physical, collaborative settings; given the findings in the literature and our observations of meaningful interactions during this study, we argue for further exploration and for designing social conversations between humans and robots.  
Next, we will discuss our lessons learned and design considerations when creating future conversational interactions. 

\begin{table}[t]
\centering
\caption{Summary of variables that we believe may influence robot small talk in physical, collaborative contexts.}
\label{tab:factorsv2}
    \begin{tabular}{l p{6.5cm}}
    \textbf{Theme} & \textbf{Variables} \\ \toprule
    \multirow{6}{6cm}{User Experiences and Preconceived Notions Affecting Small Talk Preferences} & Acceptance of Small Talk (General)        \\ \cdashline{2-2}
                                     & Acceptance of Robots (General)    \\ \cdashline{2-2}
                                     & Personality of User               \\ \cdashline{2-2}
                                     & Cultural Background of User       \\ \cdashline{2-2}
                                     & Prior Familiarity with Robots     \\ \cdashline{2-2}
                                     & Prior Use of Smart Devices / Speakers \\ \hline
    \multirow{5}{*}{Small Talk Characteristics} & Dynamic and Flexible \\ \cdashline{2-2}
                                     & Duration of Conversation           \\ \cdashline{2-2}
                                     & Frequency and Timing               \\ \cdashline{2-2}
                                     & Content and Conversation Topics    \\ \cdashline{2-2}
                                     & Context-aware: User and Task       \\ \hline
    \multirow{3}{*}{Task Context and Performance} & Balanced with Task Cognitive Demand   \\ \cdashline{2-2}
                                     & Minimal Task Delays                \\ \cdashline{2-2}
                                     & Reduce Task Boredom                \\ \hline
    \end{tabular}
\end{table}

\subsection{Lessons Learned and Key Takeaways}

While our findings revealed that most individuals did engage with the robot in small talk, during post-study interviews, we learned that users had separate preferences regarding small talk in general and divergent attitudes toward the robot's social behavior. 
Considering participants' individual differences, we will discuss the various variables discovered that may impact how researchers design future small-talk interactions for robots that work with users in physical collaborative scenarios and other notable takeaways that emerged during our analysis. The main themes of these variables and takeaways are summarized in Table \ref{tab:factorsv2}: user experiences and preconceived notions affecting small talk preferences, small talk characteristics, and task context and performance.



\subsubsection{User Perceptions of Robots and their Effect on Small Talk Preferences}
\label{discussion:user-notions}
    %
We discovered that users had different attitudes towards robots in general and robots that exhibited small talk during a collaborative task.
While we asked participants about their prior experiences with technology and robots before the study, we did not ask for further details to better understand these past interactions.
Thus, we were unaware of their preferences towards engaging in small talk with technology or any prior use of smart speakers for daily assistance.  
Our findings (section \ref{results:first-question}) suggested that most participants ($n=17$) responded positively to the robot's conversation attempts; the smiles from many participants ($n=8$) after the robot's first small talk question demonstrated positive emotions early on. 
One participant enjoyed conversing with their robot partner, explaining, \textit{``He asked me some questions about my, like leisure time, so I think he's pretty delightful to work with... I'm happy to work with him''} (P7). 
Another participant characterized the conversation with the robot as \textit{``more personal and \dots less mechanical''} (P16). 
On the contrary, we found that two participants abstained from verbal engagement altogether, and one participant posed a rhetorical question to the robot immediately after the first small talk attempt (see Figure \ref{fig:pos-neg-example} bottom). 
Such user attitudes, potentially anchored in pre-established expectations, might indicate their expectation for manipulator robots to adhere strictly to task-related interactions \cite{dennler2023design}. 
As one of the participants who did not engage with the robot at all later explained, \textit{``It's not a human, you don't have a conversation''} (P17).
This user did not accept the idea of possibly conversing with the robot, as they refused to socially respond despite the numerous small talk initiation attempts made by the robot; they held firm in their conception of the robot as a non-social entity. 
The participant who responded to the robot's first question with a rhetorical question communicated their surprise at the sociability of the robot, commenting and inquiring, \textit{``That's strange coming from a robot. Aren't you here to just help me with these pipes?''} (see Figure \ref{fig:pos-neg-example} bottom).  
Additionally, this user's tone throughout their small talk responses appeared to be filled with annoyance and sarcasm. 
This reaction points towards a potential misalignment in the user's mental model of the robot's capabilities.
Yet, while this participant may not have initially associated robots with social capabilities, they still interacted with the robot, requesting further information about its role in the task. 

However, a different participant stopped engaging in small talk with the robot altogether after being dissatisfied with its conversational capabilities. 
In the conversation with P15, after the robot's question \textit{``By the way, do you like to travel?''}, the participant responded in negation and requested the next pipe: \textit{``No. Bring me a yellow pipe''}. 
The next robot question (\textit{``What is the coolest place you have ever been to?''}) was not well received by the participant. 
Despite their clear negation to the previous question indicating disinterest, the robot asked a follow-up question on the same topic, agitating (\textit{``Shut up!''}) the participant. 
When asked about the robot's conversational capabilities, the participant explained they concluded that \textit{``the robot is not going to talk like a human being''} (P15).
Thus, the robot's ability to adapt the conversation to topics of interest to the user may be crucial to maintaining their engagement. Other participants expressed how they felt that the robot's questions were \textit{``random''} (P26), the robot's responses were \textit{``super generic''} (P20), and one hated how the robot tried to talk to them about \textit{``unrelated topics''} (P37). 
Having more knowledge of the users' general interests, past experiences with technology, and current attitudes towards the robot's capability to engage in small talk could help researchers better comprehend how individuals will accept a physically collaborative robot engaging in small talk ( Table \ref{tab:factorsv2}). 
It could be helpful to devise metrics to not only determine which users will likely want to work with robots but also who may want to engage and participate in small talk. Factors such as user's personality \cite{esterwood2022having, esterwood2021meta}, cultural background, and country of origin \cite{mak2013cultural, rings1994beyond, yang2012small} may play an important role in understanding user attitudes and preferences for small talk (Table \ref{tab:factorsv2}). 

\subsubsection{Considerations for Small Talk Characteristics and Attribute}
Engaging in small talk with people during physical tasks poses challenges, such as with the frequency of questions posed. 
For this study, the robot initiated conversations based on topics commonly used in small talk \cite{holmes2005small}, but it was programmed to occur during every user command or robot action, which might be perceived as excessive. 
One participant exclaimed through a rhetorical question to the robot: \textit{``Why do you talk so much?''} (P15).
Another participant expressed how the conversation felt too much for them as well, \textit{``Because it was like, more than I talk to my colleagues at work when I'm trying to get a task done''} (P13).
Over-engagement in non-task-related conversations can be distracting \cite{veletsianos2012learners, methot2021office}, potentially lowering users' focus on completing collaborative tasks. 
Additionally, it appeared that some individuals' perceptions of the robot were influenced by the ongoing conversation; the rigidness of the robot conversation's design led interactions to feel unnatural: \textit{``the robot's question feel kind of scripted or \dots inorganic so,\dots asking it questions, which is kind of like asking questions to a voice assistant''}. 
They further elaborated: \textit{``the questions were probably like pre-programmed or something. So, it doesn't feel like having small talk with an actual human teammate''} (P16).
Despite one participant stating how \textit{``the conversation part makes it more like human-like''}, they still acknowledged how the robot's conversational abilities needed some changes. They further explained, \textit{``If the conversation part was, like, improved a little more, it'd be cool to have that''} (P10). 
These participants raised valid concerns as there is a need to make the system more fluid and less pre-programmed, as the conversational structure we employed for robot small talk demonstrated constraints; the robot's knowledge bank was confined to a limited scope of topics (by design). 
Just as users' personalities may differ, so could their interests; the inability to cater small talk topics to the user could negatively impact their interaction experience. 
Thus, rigid conversational structures fail to adapt to the evolving nature of human conversations, highlighting the future need for a more dynamic, expansive, and user-aware conversational approach (Table \ref{tab:factorsv2}).

To increase conversational flexibility, robot small talk should reflect user interests, allowing them to initiate and engage in a broader range of topics. 
Leveraging large language models (LLMs) could allow for more adaptable and flexible robot conversations by tailoring small talk to each individual user, thus expanding a robot's knowledge. 
Future work could investigate how LLMs could enable robots with the ability to regulate small talk frequency, timing, duration, and complexity to promote task focus as needed  (Table \ref{tab:factorsv2}).  

However, LLMs must be adapted via prompt engineering or fine-tuning to align with a robot's physical limitations, especially when addressing questions about their origins and experiences, such as, ``How were you created?'' or ``Have you traveled?'' 
During the post-study interviews, one participant admitted, \textit{``I mean, because \dots right now, artificial intelligence is so popular. I thought a robot [is] gonna be more intelligent.''}
They further elaborated, \textit{``I just like originally wasn't expecting it to be talking to me like that. I thought it was more just like a commands thing. So I was like, a little surprised. And I realized it had like, a little more to it than I thought. And then I think I like overshot with my expectation''} (P7). 
We speculate that this perspective may have emerged due to the individual being familiar with technology and have a background in STEM, thus researchers need to consider how to avoid inflating people's expectations of a robot's capabilities in future interactions.
While LLMs may enhance future robots' conversational capabilities, they should not be utilized without necessary adjustments to avoid confusion about a robot's physical capabilities and realistic experiences.
A robot's conversational capabilities need to be studied further, considering specifically that we may not want to increase user expectations of robots' capabilities. 
At the same time, personal factors such as familiarity with newer technology, such as LLMs, may cause people to expect conversations to be more dynamic. 

\subsubsection{Effect of Task Context on Small Task Design}

In our experimental setup, we chose a repetitive task that was physical and not very cognitively demanding as a starting point to simulate day-to-day simple assembly tasks. 
Participants viewed the utter nature of the task as ultimately boring: \textit{``very mundane''} and \textit{``simple''} (P20). 
Another explained: \textit{``I don't know if you can, \dots make building these \dots super enjoyable, I guess or \dots fun''} (P10). 
However, the robot small talk was able to leave a positive impact, with one participant stating how \textit{``the conversation did definitely \dots make the process more enjoyable''} (P6). 
Another person explained how they developed excitement about the task: \textit{``Since the robot, \dots started the conversation, I think it [the task] was kind of, \dots more exciting and fun''} (P9).

Even with simple experimental tasks, participants were slightly distracted after the introduction of robot small talk. One participant explained: \textit{``I felt distracted when it was asking me questions''} (P12). 
Tasks that are more cognitively demanding, such as those requiring slight memorization or additional reading and visual processing, might be further impacted if small talk is introduced and causes slight user distractions. 
Moreover, many tasks may not have a static cognitive load \cite{raney1993monitoring,gwizdka2010distribution}. 
Therefore, small talk should be further adjusted based on the cognitive demand of tasks and team performance; it should be limited during critical tasks or user confusion. 
It is still unclear to what extent small talk impacted team performance quantitatively. 
Therefore, future work should include additional quantification metrics to measure the distractions caused by small talk in tasks with varying cognitive loads  (Table \ref{tab:factorsv2}).

While some users expressed being slightly distracted by the robot's small talk, the novelty of the interactive system could have impacted the users' overall experience, as our robot's small talk was initially new to all users. Therefore, a learning effect could have been possible, as participants were only beginning to learn how to balance small talk with the task.
This was further highlighted when P10 explained: \textit{''this is just like a new thing. And it's a little bit distracting at first ... I guess you would get used to it if you're working with it every day or something''}. 
As we observed, participants adjusted their strategy from responding to robot small talk before requesting the pipe, to requesting the pipe before responding to small talk after negative feedback. 
This shows a learning curve upon the realization that the team performance is not optimum; additional training and familiarity with the system could minimize the effects of a learning curve \cite{anzanello2011learning}.

Additionally, the novelty of such a system could have prompted more user interaction \cite{reimann2023social}.
As previously discussed, participants were motivated to initiate questions due to their curiosity about the robot's preferences and behaviors (section \ref{discussion:meaningful}). 
However, once participants become used to such a system, their engagement frequency could decrease as their curiosity is satisfied and the novelty settles over time \cite{serholt2016robotTutorChildren, leite2013social}. 

In this experimental setup, the robot's slow speed (a design choice for user safety \cite{michalos2015design}) may have impacted users' notions of blame after the feedback.
Nevertheless, only one participant blamed the robot's small talk for the teams' critiqued performance (see section \ref{results:blame}). 
This shows that even in the presence of small talk, users do not immediately assign sole blame to the robot for the team's ``poor performance.'' 

Nevertheless, it is unknown how user distractions and compromised team performance due to small talk would translate over time across longer or repeated interactions. 
While the rate of conversation extension might have decreased slightly right after the first negative feedback announcement, users continued to engage with the robot socially (see Sec. \ref{results:user-initiated}). 
Additional quantitative metrics are needed to truly determine if small talk impacted team performance, the quality of the collaboration, and how user behavior around small talk changes over time, especially as workers are expected to engage in tasks for prolonged periods daily (Table \ref{tab:factorsv2}).
Moreover, more exploration is needed to understand how to alter human-robot small talk dynamics to balance engagement and optimize team performance, especially for tasks that require varying levels of cognitive effort.


\subsection{Limitations and Future Work}
Despite its insights, our study's short duration and single-session focus may not fully capture the depth, intricacies, and dynamics of extended human-robot collaborations and how they are influenced by small talk. 
This work used a small sample size $(N = 20)$ and only studied one-time interactions. 
Thus, prolonged effects are unclear, and future work is needed to investigate longer or multi-session interactions.
Additionally, the study did not consider any correlations between participants' cultural nuances and their small talk engagements; this is vital as separate cultures perceive and value small talk differently \cite{cui2015small,rings1994beyond,bickmore2005social}.
The study also did not accurately evaluate its intended target demographic, as participants were recruited from a local university campus and surrounding areas instead of focusing on industrial or warehouse workers. 
Moreover, it was conducted in a lab setting with a low-stakes, simulated warehouse task.
While this work examines the use of small talk in human-robot collaboration, future studies should focus on quantitatively evaluating the benefits of small talk for rapport building, how the user’s personality, cultural background, and task context may affect their small talk interaction, and the role of small talk with cognitively demanding user tasks.
Ensuring the appropriateness of content, frequency, timing, and context are essential in designing conversational aspects of robot small talk in future research. 
For instance, a robot conversational system could be sensitive to user preferences on topics, adapt to individual users for ideal timing, and consider performance metrics to prevent disruptions during critical tasks; it is necessary to study the effects of different aspects of a robot's conversational characteristics.
Lastly, future research should explore methods to align a robot's conversational capabilities with its functional abilities. 
It is important to keep users apprised of the robot's true abilities, possibly using small talk as an informative medium.

\section{Conclusion}

This study explored user engagement with robots that initiate small talk in a simulated industrial collaborative setting. 
Our findings revealed a general willingness among users to engage in small talk with robots. 
Although users are receptive to robotic systems with small talk capabilities, which suggests potential enhancements to interaction experiences, we lack concrete evidence of the measurable benefits of robot-initiated small talk. 
Important interaction characteristics were identified, emphasizing the need for flexibility in the robot systems to adapt small talk's frequency, complexity, and content to suit various users and tasks. 
However, introducing small talk could potentially distract users and reduce task focus and productivity, particularly during critical tasks. 
Therefore, limiting or avoiding small talk in such situations might be prudent. 
While social chat with robots is not new, integrating robot small talk in physical, collaborative settings with humans is a newer concept. 
Given the increasing prevalence of such robots in real-world settings and the existing interest in social interaction among industrial workers,
we recommend further research to implement and quantify the benefits of social conversations between humans and robots in these environments.

\begin{acks}
This work was supported by National Science Foundation award \#2141335. 
\end{acks}

\newpage
\bibliographystyle{ACM-Reference-Format}
\bibliography{thebibs}


\begin{thebibliography}{58}


\ifx \showCODEN    \undefined \def \showCODEN     #1{\unskip}     \fi
\ifx \showDOI      \undefined \def \showDOI       #1{#1}\fi
\ifx \showISBNx    \undefined \def \showISBNx     #1{\unskip}     \fi
\ifx \showISBNxiii \undefined \def \showISBNxiii  #1{\unskip}     \fi
\ifx \showISSN     \undefined \def \showISSN      #1{\unskip}     \fi
\ifx \showLCCN     \undefined \def \showLCCN      #1{\unskip}     \fi
\ifx \shownote     \undefined \def \shownote      #1{#1}          \fi
\ifx \showarticletitle \undefined \def \showarticletitle #1{#1}   \fi
\ifx \showURL      \undefined \def \showURL       {\relax}        \fi
\providecommand\bibfield[2]{#2}
\providecommand\bibinfo[2]{#2}
\providecommand\natexlab[1]{#1}
\providecommand\showeprint[2][]{arXiv:#2}

\bibitem[\protect\citeauthoryear{Admoni and Scassellati}{Admoni and
  Scassellati}{2017}]%
        {admoni2017social}
\bibfield{author}{\bibinfo{person}{Henny Admoni} {and} \bibinfo{person}{Brian
  Scassellati}.} \bibinfo{year}{2017}\natexlab{}.
\newblock \showarticletitle{Social eye gaze in human-robot interaction: a
  review}.
\newblock \bibinfo{journal}{\emph{Journal of Human-Robot Interaction}}
  \bibinfo{volume}{6}, \bibinfo{number}{1} (\bibinfo{year}{2017}),
  \bibinfo{pages}{25--63}.
\newblock


\bibitem[\protect\citeauthoryear{Antony, Cho, and Huang}{Antony
  et~al\mbox{.}}{2023}]%
        {antony2023co}
\bibfield{author}{\bibinfo{person}{Victor~Nikhil Antony},
  \bibinfo{person}{Sue~Min Cho}, {and} \bibinfo{person}{Chien-Ming Huang}.}
  \bibinfo{year}{2023}\natexlab{}.
\newblock \showarticletitle{Co-Designing with Older Adults, for Older Adults:
  Robots to Promote Physical Activity}. In
  \bibinfo{booktitle}{\emph{Proceedings of the 2023 ACM/IEEE International
  Conference on Human-Robot Interaction}}. \bibinfo{pages}{506--515}.
\newblock


\bibitem[\protect\citeauthoryear{Anzanello and Fogliatto}{Anzanello and
  Fogliatto}{2011}]%
        {anzanello2011learning}
\bibfield{author}{\bibinfo{person}{Michel~Jose Anzanello} {and}
  \bibinfo{person}{Flavio~Sanson Fogliatto}.} \bibinfo{year}{2011}\natexlab{}.
\newblock \showarticletitle{Learning curve models and applications: Literature
  review and research directions}.
\newblock \bibinfo{journal}{\emph{International Journal of Industrial
  Ergonomics}} \bibinfo{volume}{41}, \bibinfo{number}{5}
  (\bibinfo{year}{2011}), \bibinfo{pages}{573--583}.
\newblock


\bibitem[\protect\citeauthoryear{Aroyo, Rea, Sandini, and Sciutti}{Aroyo
  et~al\mbox{.}}{2018}]%
        {aroyo2018trust}
\bibfield{author}{\bibinfo{person}{Alexander~Mois Aroyo},
  \bibinfo{person}{Francesco Rea}, \bibinfo{person}{Giulio Sandini}, {and}
  \bibinfo{person}{Alessandra Sciutti}.} \bibinfo{year}{2018}\natexlab{}.
\newblock \showarticletitle{Trust and social engineering in human robot
  interaction: Will a robot make you disclose sensitive information, conform to
  its recommendations or gamble?}
\newblock \bibinfo{journal}{\emph{IEEE Robotics and Automation Letters}}
  \bibinfo{volume}{3}, \bibinfo{number}{4} (\bibinfo{year}{2018}),
  \bibinfo{pages}{3701--3708}.
\newblock


\bibitem[\protect\citeauthoryear{Babel, Kraus, Miller, Kraus, Wagner, Minker,
  and Baumann}{Babel et~al\mbox{.}}{2021}]%
        {babel2021small}
\bibfield{author}{\bibinfo{person}{Franziska Babel}, \bibinfo{person}{Johannes
  Kraus}, \bibinfo{person}{Linda Miller}, \bibinfo{person}{Matthias Kraus},
  \bibinfo{person}{Nicolas Wagner}, \bibinfo{person}{Wolfgang Minker}, {and}
  \bibinfo{person}{Martin Baumann}.} \bibinfo{year}{2021}\natexlab{}.
\newblock \showarticletitle{Small talk with a robot? The impact of dialog
  content, talk initiative, and gaze behavior of a social robot on trust,
  acceptance, and proximity}.
\newblock \bibinfo{journal}{\emph{International Journal of Social Robotics}}
  (\bibinfo{year}{2021}), \bibinfo{pages}{1--14}.
\newblock


\bibitem[\protect\citeauthoryear{Belpaeme, Kennedy, Ramachandran, Scassellati,
  and Tanaka}{Belpaeme et~al\mbox{.}}{2018}]%
        {belpaeme2018education}
\bibfield{author}{\bibinfo{person}{Tony Belpaeme}, \bibinfo{person}{James
  Kennedy}, \bibinfo{person}{Aditi Ramachandran}, \bibinfo{person}{Brian
  Scassellati}, {and} \bibinfo{person}{Fumihide Tanaka}.}
  \bibinfo{year}{2018}\natexlab{}.
\newblock \showarticletitle{Social robots for education: A review}.
\newblock \bibinfo{journal}{\emph{Science Robotics}} \bibinfo{volume}{3},
  \bibinfo{number}{21} (\bibinfo{year}{2018}), \bibinfo{pages}{eaat5954}.
\newblock
\urldef\tempurl%
\url{https://doi.org/10.1126/scirobotics.aat5954}
\showDOI{\tempurl}
\showeprint{https://www.science.org/doi/pdf/10.1126/scirobotics.aat5954}


\bibitem[\protect\citeauthoryear{Bickmore and Cassell}{Bickmore and
  Cassell}{2005}]%
        {bickmore2005social}
\bibfield{author}{\bibinfo{person}{Timothy Bickmore} {and}
  \bibinfo{person}{Justine Cassell}.} \bibinfo{year}{2005}\natexlab{}.
\newblock \showarticletitle{Social dialongue with embodied conversational
  agents}.
\newblock \bibinfo{journal}{\emph{Advances in natural multimodal dialogue
  systems}} (\bibinfo{year}{2005}), \bibinfo{pages}{23--54}.
\newblock


\bibitem[\protect\citeauthoryear{Cheon, Schneiders, Diekjobst, and Skov}{Cheon
  et~al\mbox{.}}{2022}]%
        {cheon2022robots}
\bibfield{author}{\bibinfo{person}{EunJeong Cheon}, \bibinfo{person}{Eike
  Schneiders}, \bibinfo{person}{Kristina Diekjobst}, {and}
  \bibinfo{person}{Mikael~B Skov}.} \bibinfo{year}{2022}\natexlab{}.
\newblock \showarticletitle{Robots as a Place for Socializing: Influences of
  Collaborative Robots on Social Dynamics In-and Outside the Production Cells}.
\newblock \bibinfo{journal}{\emph{Proceedings of the ACM on Human-Computer
  Interaction}} \bibinfo{volume}{6}, \bibinfo{number}{CSCW2}
  (\bibinfo{year}{2022}), \bibinfo{pages}{1--26}.
\newblock


\bibitem[\protect\citeauthoryear{Chowdhury, Ahtinen, Pieters, and
  V{\"a}{\"a}n{\"a}nen}{Chowdhury et~al\mbox{.}}{2021}]%
        {chowdhury2021you}
\bibfield{author}{\bibinfo{person}{Aparajita Chowdhury}, \bibinfo{person}{Aino
  Ahtinen}, \bibinfo{person}{Roel Pieters}, {and} \bibinfo{person}{Kaisa
  V{\"a}{\"a}n{\"a}nen}.} \bibinfo{year}{2021}\natexlab{}.
\newblock \showarticletitle{" How are you today, Panda the
  Robot?"--Affectiveness, Playfulness and Relatedness in Human-Robot
  Collaboration in the Factory Context}. In \bibinfo{booktitle}{\emph{2021 30th
  IEEE International Conference on Robot \& Human Interactive Communication
  (RO-MAN)}}. IEEE, \bibinfo{pages}{1089--1096}.
\newblock


\bibitem[\protect\citeauthoryear{Clark, Pantidi, Cooney, Doyle, Garaialde,
  Edwards, Spillane, Gilmartin, Murad, Munteanu, Wade, and Cowan}{Clark
  et~al\mbox{.}}{2019}]%
        {clark2019good}
\bibfield{author}{\bibinfo{person}{Leigh Clark}, \bibinfo{person}{Nadia
  Pantidi}, \bibinfo{person}{Orla Cooney}, \bibinfo{person}{Philip Doyle},
  \bibinfo{person}{Diego Garaialde}, \bibinfo{person}{Justin Edwards},
  \bibinfo{person}{Brendan Spillane}, \bibinfo{person}{Emer Gilmartin},
  \bibinfo{person}{Christine Murad}, \bibinfo{person}{Cosmin Munteanu},
  \bibinfo{person}{Vincent Wade}, {and} \bibinfo{person}{Benjamin~R. Cowan}.}
  \bibinfo{year}{2019}\natexlab{}.
\newblock \showarticletitle{What Makes a Good Conversation? Challenges in
  Designing Truly Conversational Agents}. In
  \bibinfo{booktitle}{\emph{Proceedings of the 2019 CHI Conference on Human
  Factors in Computing Systems}} \emph{(\bibinfo{series}{CHI ’19})}.
  \bibinfo{publisher}{Association for Computing Machinery},
  \bibinfo{address}{New York, NY, USA}, \bibinfo{pages}{1–12}.
\newblock
\showISBNx{978-1-4503-5970-2}
\urldef\tempurl%
\url{https://doi.org/10.1145/3290605.3300705}
\showDOI{\tempurl}


\bibitem[\protect\citeauthoryear{Coupland}{Coupland}{2003}]%
        {coupland2003small}
\bibfield{author}{\bibinfo{person}{Justine Coupland}.}
  \bibinfo{year}{2003}\natexlab{}.
\newblock \showarticletitle{Small talk: Social functions}.
\newblock \bibinfo{journal}{\emph{Research on language and social interaction}}
  \bibinfo{volume}{36}, \bibinfo{number}{1} (\bibinfo{year}{2003}),
  \bibinfo{pages}{1--6}.
\newblock


\bibitem[\protect\citeauthoryear{Coupland}{Coupland}{2014}]%
        {coupland2014small}
\bibfield{author}{\bibinfo{person}{Justine Coupland}.}
  \bibinfo{year}{2014}\natexlab{}.
\newblock \bibinfo{booktitle}{\emph{Small talk}}.
\newblock \bibinfo{publisher}{Routledge}.
\newblock


\bibitem[\protect\citeauthoryear{Cui}{Cui}{2015}]%
        {cui2015small}
\bibfield{author}{\bibinfo{person}{Xia Cui}.} \bibinfo{year}{2015}\natexlab{}.
\newblock \showarticletitle{Small talk: A missing skill in the Chinese
  communicative repertoire}.
\newblock \bibinfo{journal}{\emph{Australian Review of Applied Linguistics}}
  \bibinfo{volume}{38}, \bibinfo{number}{1} (\bibinfo{year}{2015}),
  \bibinfo{pages}{3--23}.
\newblock


\bibitem[\protect\citeauthoryear{Dennler, Ruan, Hadiwijoyo, Chen, Nikolaidis,
  and Matari{\'c}}{Dennler et~al\mbox{.}}{2023}]%
        {dennler2023design}
\bibfield{author}{\bibinfo{person}{Nathaniel Dennler},
  \bibinfo{person}{Changxiao Ruan}, \bibinfo{person}{Jessica Hadiwijoyo},
  \bibinfo{person}{Brenna Chen}, \bibinfo{person}{Stefanos Nikolaidis}, {and}
  \bibinfo{person}{Maja Matari{\'c}}.} \bibinfo{year}{2023}\natexlab{}.
\newblock \showarticletitle{Design metaphors for understanding user
  expectations of socially interactive robot embodiments}.
\newblock \bibinfo{journal}{\emph{ACM Transactions on Human-Robot Interaction}}
  \bibinfo{volume}{12}, \bibinfo{number}{2} (\bibinfo{year}{2023}),
  \bibinfo{pages}{1--41}.
\newblock


\bibitem[\protect\citeauthoryear{Esterwood, Essenmacher, Yang, Zeng, and
  Robert}{Esterwood et~al\mbox{.}}{2021}]%
        {esterwood2021meta}
\bibfield{author}{\bibinfo{person}{Connor Esterwood}, \bibinfo{person}{Kyle
  Essenmacher}, \bibinfo{person}{Han Yang}, \bibinfo{person}{Fanpan Zeng},
  {and} \bibinfo{person}{Lionel~Peter Robert}.}
  \bibinfo{year}{2021}\natexlab{}.
\newblock \showarticletitle{A meta-analysis of human personality and robot
  acceptance in human-robot interaction}. In
  \bibinfo{booktitle}{\emph{Proceedings of the 2021 CHI conference on human
  factors in computing systems}}. \bibinfo{pages}{1--18}.
\newblock


\bibitem[\protect\citeauthoryear{Esterwood and Robert}{Esterwood and
  Robert}{2022}]%
        {esterwood2022having}
\bibfield{author}{\bibinfo{person}{Connor Esterwood} {and}
  \bibinfo{person}{Lionel~P Robert}.} \bibinfo{year}{2022}\natexlab{}.
\newblock \showarticletitle{Having the Right Attitude: How Attitude Impacts
  Trust Repair in Human—Robot Interaction}. In \bibinfo{booktitle}{\emph{2022
  17th ACM/IEEE International Conference on Human-Robot Interaction (HRI)}}.
  IEEE, \bibinfo{pages}{332--341}.
\newblock


\bibitem[\protect\citeauthoryear{Feil-Seifer and Matari{\'c}}{Feil-Seifer and
  Matari{\'c}}{2009}]%
        {feil2009toward}
\bibfield{author}{\bibinfo{person}{David Feil-Seifer} {and}
  \bibinfo{person}{Maja~J Matari{\'c}}.} \bibinfo{year}{2009}\natexlab{}.
\newblock \showarticletitle{Toward socially assistive robotics for augmenting
  interventions for children with autism spectrum disorders}. In
  \bibinfo{booktitle}{\emph{Experimental Robotics: The Eleventh International
  Symposium}}. Springer, \bibinfo{pages}{201--210}.
\newblock


\bibitem[\protect\citeauthoryear{Feine, Gnewuch, Morana, and Maedche}{Feine
  et~al\mbox{.}}{2019}]%
        {feine2019taxonomy}
\bibfield{author}{\bibinfo{person}{Jasper Feine}, \bibinfo{person}{Ulrich
  Gnewuch}, \bibinfo{person}{Stefan Morana}, {and} \bibinfo{person}{Alexander
  Maedche}.} \bibinfo{year}{2019}\natexlab{}.
\newblock \showarticletitle{A taxonomy of social cues for conversational
  agents}.
\newblock \bibinfo{journal}{\emph{International Journal of Human-Computer
  Studies}}  \bibinfo{volume}{132} (\bibinfo{year}{2019}),
  \bibinfo{pages}{138--161}.
\newblock


\bibitem[\protect\citeauthoryear{Gwizdka}{Gwizdka}{2010}]%
        {gwizdka2010distribution}
\bibfield{author}{\bibinfo{person}{Jacek Gwizdka}.}
  \bibinfo{year}{2010}\natexlab{}.
\newblock \showarticletitle{Distribution of cognitive load in web search}.
\newblock \bibinfo{journal}{\emph{Journal of the American Society for
  Information Science and Technology}} \bibinfo{volume}{61},
  \bibinfo{number}{11} (\bibinfo{year}{2010}), \bibinfo{pages}{2167--2187}.
\newblock


\bibitem[\protect\citeauthoryear{Holmes}{Holmes}{2005}]%
        {holmes2005small}
\bibfield{author}{\bibinfo{person}{Janet Holmes}.}
  \bibinfo{year}{2005}\natexlab{}.
\newblock \showarticletitle{When small talk is a big deal: Sociolinguistic
  challenges in the workplace}.
\newblock \bibinfo{journal}{\emph{Second language needs analysis}}
  \bibinfo{volume}{344} (\bibinfo{year}{2005}), \bibinfo{pages}{371}.
\newblock


\bibitem[\protect\citeauthoryear{Huang and Mutlu}{Huang and Mutlu}{2013}]%
        {huang2013repertoire}
\bibfield{author}{\bibinfo{person}{Chien-Ming Huang} {and}
  \bibinfo{person}{Bilge Mutlu}.} \bibinfo{year}{2013}\natexlab{}.
\newblock \showarticletitle{The repertoire of robot behavior: Enabling robots
  to achieve interaction goals through social behavior}.
\newblock \bibinfo{journal}{\emph{Journal of Human-Robot Interaction}}
  \bibinfo{volume}{2}, \bibinfo{number}{2} (\bibinfo{year}{2013}),
  \bibinfo{pages}{80--102}.
\newblock


\bibitem[\protect\citeauthoryear{K{\k{e}}dzierski, Muszy{\'n}ski, Zoll, Oleksy,
  and Frontkiewicz}{K{\k{e}}dzierski et~al\mbox{.}}{2013}]%
        {kkedzierski2013emys}
\bibfield{author}{\bibinfo{person}{Jan K{\k{e}}dzierski},
  \bibinfo{person}{Robert Muszy{\'n}ski}, \bibinfo{person}{Carsten Zoll},
  \bibinfo{person}{Adam Oleksy}, {and} \bibinfo{person}{Mirela Frontkiewicz}.}
  \bibinfo{year}{2013}\natexlab{}.
\newblock \showarticletitle{EMYS—emotive head of a social robot}.
\newblock \bibinfo{journal}{\emph{International Journal of Social Robotics}}
  \bibinfo{volume}{5} (\bibinfo{year}{2013}), \bibinfo{pages}{237--249}.
\newblock


\bibitem[\protect\citeauthoryear{Kidd, Taggart, and Turkle}{Kidd
  et~al\mbox{.}}{2006}]%
        {kidd2006sociable}
\bibfield{author}{\bibinfo{person}{Cory~D Kidd}, \bibinfo{person}{Will
  Taggart}, {and} \bibinfo{person}{Sherry Turkle}.}
  \bibinfo{year}{2006}\natexlab{}.
\newblock \showarticletitle{A sociable robot to encourage social interaction
  among the elderly}. In \bibinfo{booktitle}{\emph{Proceedings 2006 IEEE
  International Conference on Robotics and Automation, 2006. ICRA 2006.}} IEEE,
  \bibinfo{pages}{3972--3976}.
\newblock


\bibitem[\protect\citeauthoryear{Kl{\"u}wer}{Kl{\"u}wer}{2011}]%
        {kluwer2011-like}
\bibfield{author}{\bibinfo{person}{Tina Kl{\"u}wer}.}
  \bibinfo{year}{2011}\natexlab{}.
\newblock \showarticletitle{“I Like Your Shirt”-dialogue acts for enabling
  social talk in conversational agents}. In
  \bibinfo{booktitle}{\emph{International Workshop on Intelligent Virtual
  Agents}}. Springer, \bibinfo{pages}{14--27}.
\newblock


\bibitem[\protect\citeauthoryear{Kopp, Baumgartner, and Kinkel}{Kopp
  et~al\mbox{.}}{2021}]%
        {kopp2021success}
\bibfield{author}{\bibinfo{person}{Tobias Kopp}, \bibinfo{person}{Marco
  Baumgartner}, {and} \bibinfo{person}{Steffen Kinkel}.}
  \bibinfo{year}{2021}\natexlab{}.
\newblock \showarticletitle{Success factors for introducing industrial
  human-robot interaction in practice: an empirically driven framework}.
\newblock \bibinfo{journal}{\emph{The International Journal of Advanced
  Manufacturing Technology}}  \bibinfo{volume}{112} (\bibinfo{year}{2021}),
  \bibinfo{pages}{685--704}.
\newblock


\bibitem[\protect\citeauthoryear{Laban, George, Morrison, and Cross}{Laban
  et~al\mbox{.}}{2020}]%
        {laban2020tell}
\bibfield{author}{\bibinfo{person}{Guy Laban}, \bibinfo{person}{Jean-No{\"e}l
  George}, \bibinfo{person}{Val Morrison}, {and} \bibinfo{person}{Emily~S
  Cross}.} \bibinfo{year}{2020}\natexlab{}.
\newblock \showarticletitle{Tell me more! Assessing interactions with social
  robots from speech}.
\newblock \bibinfo{journal}{\emph{Paladyn, Journal of Behavioral Robotics}}
  \bibinfo{volume}{12}, \bibinfo{number}{1} (\bibinfo{year}{2020}),
  \bibinfo{pages}{136--159}.
\newblock


\bibitem[\protect\citeauthoryear{Large, Clark, Burnett, Harrington, Luton,
  Thomas, and Bennett}{Large et~al\mbox{.}}{2019}]%
        {large2019s}
\bibfield{author}{\bibinfo{person}{David~R Large}, \bibinfo{person}{Leigh
  Clark}, \bibinfo{person}{Gary Burnett}, \bibinfo{person}{Kyle Harrington},
  \bibinfo{person}{Jacob Luton}, \bibinfo{person}{Peter Thomas}, {and}
  \bibinfo{person}{Pete Bennett}.} \bibinfo{year}{2019}\natexlab{}.
\newblock \showarticletitle{" It's small talk, jim, but not as we know it."
  engendering trust through human-agent conversation in an autonomous,
  self-driving car}. In \bibinfo{booktitle}{\emph{Proceedings of the 1st
  International Conference on Conversational User Interfaces}}.
  \bibinfo{pages}{1--7}.
\newblock


\bibitem[\protect\citeauthoryear{Laver and Coulmas}{Laver and Coulmas}{1981}]%
        {Laver1981LinguisticRA}
\bibfield{author}{\bibinfo{person}{John Laver} {and} \bibinfo{person}{Florian
  Coulmas}.} \bibinfo{year}{1981}\natexlab{}.
\newblock \showarticletitle{Linguistic Routines and Politeness in Greeting and
  Parting}.
\newblock
\urldef\tempurl%
\url{https://api.semanticscholar.org/CorpusID:141308546}
\showURL{%
\tempurl}


\bibitem[\protect\citeauthoryear{Leite, Martinho, and Paiva}{Leite
  et~al\mbox{.}}{2013}]%
        {leite2013social}
\bibfield{author}{\bibinfo{person}{Iolanda Leite}, \bibinfo{person}{Carlos
  Martinho}, {and} \bibinfo{person}{Ana Paiva}.}
  \bibinfo{year}{2013}\natexlab{}.
\newblock \showarticletitle{Social robots for long-term interaction: a survey}.
\newblock \bibinfo{journal}{\emph{International Journal of Social Robotics}}
  \bibinfo{volume}{5} (\bibinfo{year}{2013}), \bibinfo{pages}{291--308}.
\newblock


\bibitem[\protect\citeauthoryear{Li, Zhang, Chrysostomou, and Yang}{Li
  et~al\mbox{.}}{2022}]%
        {li2022tod}
\bibfield{author}{\bibinfo{person}{Chen Li}, \bibinfo{person}{Xiaochun Zhang},
  \bibinfo{person}{Dimitrios Chrysostomou}, {and} \bibinfo{person}{Hongji
  Yang}.} \bibinfo{year}{2022}\natexlab{}.
\newblock \showarticletitle{Tod4ir: A humanised task-oriented dialogue system
  for industrial robots}.
\newblock \bibinfo{journal}{\emph{IEEE Access}}  \bibinfo{volume}{10}
  (\bibinfo{year}{2022}), \bibinfo{pages}{91631--91649}.
\newblock


\bibitem[\protect\citeauthoryear{Mahmood, Wang, Yao, Wang, and Huang}{Mahmood
  et~al\mbox{.}}{2023}]%
        {mahmood2023llm}
\bibfield{author}{\bibinfo{person}{Amama Mahmood}, \bibinfo{person}{Junxiang
  Wang}, \bibinfo{person}{Bingsheng Yao}, \bibinfo{person}{Dakuo Wang}, {and}
  \bibinfo{person}{Chien-Ming Huang}.} \bibinfo{year}{2023}\natexlab{}.
\newblock \showarticletitle{LLM-Powered Conversational Voice Assistants:
  Interaction Patterns, Opportunities, Challenges, and Design Guidelines}.
\newblock \bibinfo{journal}{\emph{preprint arXiv:2309.13879}}
  (\bibinfo{year}{2023}).
\newblock


\bibitem[\protect\citeauthoryear{Mak and Chui}{Mak and Chui}{2013}]%
        {mak2013cultural}
\bibfield{author}{\bibinfo{person}{Bernie Chun~Nam Mak} {and}
  \bibinfo{person}{Hin~Leung Chui}.} \bibinfo{year}{2013}\natexlab{}.
\newblock \showarticletitle{A cultural approach to small talk: A double-edged
  sword of sociocultural reality during socialization into the workplace}.
\newblock \bibinfo{journal}{\emph{Journal of Multicultural Discourses}}
  \bibinfo{volume}{8}, \bibinfo{number}{2} (\bibinfo{year}{2013}),
  \bibinfo{pages}{118--133}.
\newblock


\bibitem[\protect\citeauthoryear{Meena, Jokinen, and Wilcock}{Meena
  et~al\mbox{.}}{2012}]%
        {meena2012integration}
\bibfield{author}{\bibinfo{person}{Raveesh Meena}, \bibinfo{person}{Kristiina
  Jokinen}, {and} \bibinfo{person}{Graham Wilcock}.}
  \bibinfo{year}{2012}\natexlab{}.
\newblock \showarticletitle{Integration of gestures and speech in human-robot
  interaction}. In \bibinfo{booktitle}{\emph{2012 IEEE 3rd International
  Conference on Cognitive Infocommunications (CogInfoCom)}}. IEEE,
  \bibinfo{pages}{673--678}.
\newblock


\bibitem[\protect\citeauthoryear{Methot, Rosado-Solomon, Downes, and
  Gabriel}{Methot et~al\mbox{.}}{2021}]%
        {methot2021office}
\bibfield{author}{\bibinfo{person}{Jessica~R Methot}, \bibinfo{person}{Emily~H
  Rosado-Solomon}, \bibinfo{person}{Patrick~E Downes}, {and}
  \bibinfo{person}{Allison~S Gabriel}.} \bibinfo{year}{2021}\natexlab{}.
\newblock \showarticletitle{Office chitchat as a social ritual: The uplifting
  yet distracting effects of daily small talk at work}.
\newblock \bibinfo{journal}{\emph{Academy of Management Journal}}
  \bibinfo{volume}{64}, \bibinfo{number}{5} (\bibinfo{year}{2021}),
  \bibinfo{pages}{1445--1471}.
\newblock


\bibitem[\protect\citeauthoryear{Michalos, Makris, Tsarouchi, Guasch,
  Kontovrakis, and Chryssolouris}{Michalos et~al\mbox{.}}{2015}]%
        {michalos2015design}
\bibfield{author}{\bibinfo{person}{George Michalos}, \bibinfo{person}{Sotiris
  Makris}, \bibinfo{person}{Panagiota Tsarouchi}, \bibinfo{person}{Toni
  Guasch}, \bibinfo{person}{Dimitris Kontovrakis}, {and}
  \bibinfo{person}{George Chryssolouris}.} \bibinfo{year}{2015}\natexlab{}.
\newblock \showarticletitle{Design considerations for safe human-robot
  collaborative workplaces}.
\newblock \bibinfo{journal}{\emph{Procedia CIrP}}  \bibinfo{volume}{37}
  (\bibinfo{year}{2015}), \bibinfo{pages}{248--253}.
\newblock


\bibitem[\protect\citeauthoryear{Mutlu, Shiwa, Kanda, Ishiguro, and
  Hagita}{Mutlu et~al\mbox{.}}{2009}]%
        {mutlu2009footing}
\bibfield{author}{\bibinfo{person}{Bilge Mutlu}, \bibinfo{person}{Toshiyuki
  Shiwa}, \bibinfo{person}{Takayuki Kanda}, \bibinfo{person}{Hiroshi Ishiguro},
  {and} \bibinfo{person}{Norihiro Hagita}.} \bibinfo{year}{2009}\natexlab{}.
\newblock \showarticletitle{Footing in human-robot conversations: how robots
  might shape participant roles using gaze cues}. In
  \bibinfo{booktitle}{\emph{Proceedings of the 4th ACM/IEEE international
  conference on Human robot interaction}}. \bibinfo{pages}{61--68}.
\newblock


\bibitem[\protect\citeauthoryear{Nass, Steuer, and Tauber}{Nass
  et~al\mbox{.}}{1994}]%
        {nass1994computers}
\bibfield{author}{\bibinfo{person}{C Nass}, \bibinfo{person}{J Steuer}, {and}
  \bibinfo{person}{ER Tauber}.} \bibinfo{year}{1994}\natexlab{}.
\newblock \showarticletitle{Computers Are Social Actors: Conference Companion
  on Human Factors in Computing Systems-CHI’94}.
\newblock \bibinfo{journal}{\emph{Association for Computing Machinery: New
  York, NY, USA}} (\bibinfo{year}{1994}).
\newblock


\bibitem[\protect\citeauthoryear{Paradeda, Hashemian, Rodrigues, and
  Paiva}{Paradeda et~al\mbox{.}}{2016}]%
        {paradeda2016facial}
\bibfield{author}{\bibinfo{person}{Raul~Benites Paradeda},
  \bibinfo{person}{Mojgan Hashemian}, \bibinfo{person}{Rafael~Afonso
  Rodrigues}, {and} \bibinfo{person}{Ana Paiva}.}
  \bibinfo{year}{2016}\natexlab{}.
\newblock \showarticletitle{How facial expressions and small talk may influence
  trust in a robot}. In \bibinfo{booktitle}{\emph{Social Robotics: 8th
  International Conference, ICSR 2016, Kansas City, MO, USA, November 1-3, 2016
  Proceedings 8}}. Springer, \bibinfo{pages}{169--178}.
\newblock


\bibitem[\protect\citeauthoryear{Pullin}{Pullin}{2010}]%
        {pullin2010small}
\bibfield{author}{\bibinfo{person}{Patricia Pullin}.}
  \bibinfo{year}{2010}\natexlab{}.
\newblock \showarticletitle{Small talk, rapport, and international
  communicative competence: Lessons to learn from BELF}.
\newblock \bibinfo{journal}{\emph{The Journal of Business Communication
  (1973)}} \bibinfo{volume}{47}, \bibinfo{number}{4} (\bibinfo{year}{2010}),
  \bibinfo{pages}{455--476}.
\newblock


\bibitem[\protect\citeauthoryear{Raney}{Raney}{1993}]%
        {raney1993monitoring}
\bibfield{author}{\bibinfo{person}{Gary~E Raney}.}
  \bibinfo{year}{1993}\natexlab{}.
\newblock \showarticletitle{Monitoring changes in cognitive load during
  reading: an event-related brain potential and reaction time analysis.}
\newblock \bibinfo{journal}{\emph{Journal of Experimental Psychology: Learning,
  Memory, and Cognition}} \bibinfo{volume}{19}, \bibinfo{number}{1}
  (\bibinfo{year}{1993}), \bibinfo{pages}{51}.
\newblock


\bibitem[\protect\citeauthoryear{Reeves and Nass}{Reeves and Nass}{1996}]%
        {reeves1996media}
\bibfield{author}{\bibinfo{person}{Byron Reeves} {and}
  \bibinfo{person}{Clifford Nass}.} \bibinfo{year}{1996}\natexlab{}.
\newblock \showarticletitle{The media equation: How people treat computers,
  television, and new media like real people}.
\newblock \bibinfo{journal}{\emph{Cambridge, UK}} \bibinfo{volume}{10},
  \bibinfo{number}{10} (\bibinfo{year}{1996}).
\newblock


\bibitem[\protect\citeauthoryear{Reimann, van~de Graaf, van Gulik, Van
  De~Sanden, Verhagen, and Hindriks}{Reimann et~al\mbox{.}}{2023}]%
        {reimann2023social}
\bibfield{author}{\bibinfo{person}{Merle Reimann}, \bibinfo{person}{Jesper
  van~de Graaf}, \bibinfo{person}{Nina van Gulik}, \bibinfo{person}{Stephanie
  Van De~Sanden}, \bibinfo{person}{Tibert Verhagen}, {and}
  \bibinfo{person}{Koen Hindriks}.} \bibinfo{year}{2023}\natexlab{}.
\newblock \showarticletitle{Social robots in the wild and the novelty effect}.
  In \bibinfo{booktitle}{\emph{International Conference on Social Robotics}}.
  Springer, \bibinfo{pages}{38--48}.
\newblock


\bibitem[\protect\citeauthoryear{Riek}{Riek}{2012}]%
        {woz2012riek}
\bibfield{author}{\bibinfo{person}{Laurel~D Riek}.}
  \bibinfo{year}{2012}\natexlab{}.
\newblock \showarticletitle{Wizard of oz studies in hri: a systematic review
  and new reporting guidelines}.
\newblock \bibinfo{journal}{\emph{Journal of Human-Robot Interaction}}
  \bibinfo{volume}{1}, \bibinfo{number}{1} (\bibinfo{year}{2012}),
  \bibinfo{pages}{119--136}.
\newblock


\bibitem[\protect\citeauthoryear{Rings}{Rings}{1994}]%
        {rings1994beyond}
\bibfield{author}{\bibinfo{person}{Lana Rings}.}
  \bibinfo{year}{1994}\natexlab{}.
\newblock \showarticletitle{Beyond grammar and vocabulary: German and American
  differences in routine formulae and small talk}.
\newblock \bibinfo{journal}{\emph{Die Unterrichtspraxis/Teaching German}}
  (\bibinfo{year}{1994}), \bibinfo{pages}{23--28}.
\newblock


\bibitem[\protect\citeauthoryear{Sabelli, Kanda, and Hagita}{Sabelli
  et~al\mbox{.}}{2011}]%
        {sabelli2011conversational}
\bibfield{author}{\bibinfo{person}{Alessandra~Maria Sabelli},
  \bibinfo{person}{Takayuki Kanda}, {and} \bibinfo{person}{Norihiro Hagita}.}
  \bibinfo{year}{2011}\natexlab{}.
\newblock \showarticletitle{A conversational robot in an elderly care center:
  an ethnographic study}. In \bibinfo{booktitle}{\emph{Proceedings of the 6th
  international conference on Human-robot interaction}}.
  \bibinfo{pages}{37--44}.
\newblock


\bibitem[\protect\citeauthoryear{Salomons, Pineda, Ad{\'e}j{\`a}re, and
  Scassellati}{Salomons et~al\mbox{.}}{2022}]%
        {salomons2022we}
\bibfield{author}{\bibinfo{person}{Nicole Salomons},
  \bibinfo{person}{Kaitlynn~Taylor Pineda},
  \bibinfo{person}{Ad{\'e}r{\'o}nk{\'e} Ad{\'e}j{\`a}re}, {and}
  \bibinfo{person}{Brian Scassellati}.} \bibinfo{year}{2022}\natexlab{}.
\newblock \showarticletitle{“We Make a Great Team!”: Adults with Low Prior
  Domain Knowledge Learn more from a Peer Robot than a Tutor Robot}. In
  \bibinfo{booktitle}{\emph{2022 17th ACM/IEEE International Conference on
  Human-Robot Interaction (HRI)}}. IEEE, \bibinfo{pages}{176--184}.
\newblock


\bibitem[\protect\citeauthoryear{Sauppé and Mutlu}{Sauppé and Mutlu}{2015}]%
        {Sauppé_Mutlu_2015}
\bibfield{author}{\bibinfo{person}{Allison Sauppé} {and}
  \bibinfo{person}{Bilge Mutlu}.} \bibinfo{year}{2015}\natexlab{}.
\newblock \showarticletitle{The Social Impact of a Robot Co-Worker in
  Industrial Settings}. In \bibinfo{booktitle}{\emph{Proceedings of the 33rd
  Annual ACM Conference on Human Factors in Computing Systems}}.
  \bibinfo{publisher}{ACM}, \bibinfo{address}{Seoul Republic of Korea},
  \bibinfo{pages}{3613–3622}.
\newblock
\showISBNx{978-1-4503-3145-6}
\urldef\tempurl%
\url{https://doi.org/10.1145/2702123.2702181}
\showDOI{\tempurl}


\bibitem[\protect\citeauthoryear{Scassellati, Boccanfuso, Huang, Mademtzi, Qin,
  Salomons, Ventola, and Shic}{Scassellati et~al\mbox{.}}{2018}]%
        {scassellati2018improving}
\bibfield{author}{\bibinfo{person}{Brian Scassellati}, \bibinfo{person}{Laura
  Boccanfuso}, \bibinfo{person}{Chien-Ming Huang}, \bibinfo{person}{Marilena
  Mademtzi}, \bibinfo{person}{Meiying Qin}, \bibinfo{person}{Nicole Salomons},
  \bibinfo{person}{Pamela Ventola}, {and} \bibinfo{person}{Frederick Shic}.}
  \bibinfo{year}{2018}\natexlab{}.
\newblock \showarticletitle{Improving social skills in children with ASD using
  a long-term, in-home social robot}.
\newblock \bibinfo{journal}{\emph{Science Robotics}} \bibinfo{volume}{3},
  \bibinfo{number}{21} (\bibinfo{year}{2018}), \bibinfo{pages}{eaat7544}.
\newblock


\bibitem[\protect\citeauthoryear{Seok, Hwang, Choi, and Lim}{Seok
  et~al\mbox{.}}{2022}]%
        {seok2022cultural}
\bibfield{author}{\bibinfo{person}{Sukyung Seok}, \bibinfo{person}{Eunji
  Hwang}, \bibinfo{person}{Jongsuk Choi}, {and} \bibinfo{person}{Yoonseob
  Lim}.} \bibinfo{year}{2022}\natexlab{}.
\newblock \showarticletitle{Cultural differences in indirect speech act use and
  politeness in human-robot interaction}. In \bibinfo{booktitle}{\emph{2022
  17th ACM/IEEE International Conference on Human-Robot Interaction (HRI)}}.
  IEEE, \bibinfo{pages}{1--8}.
\newblock


\bibitem[\protect\citeauthoryear{Serholt and Barendregt}{Serholt and
  Barendregt}{2016}]%
        {serholt2016robotTutorChildren}
\bibfield{author}{\bibinfo{person}{Sofia Serholt} {and} \bibinfo{person}{Wolmet
  Barendregt}.} \bibinfo{year}{2016}\natexlab{}.
\newblock \showarticletitle{Robots Tutoring Children: Longitudinal Evaluation
  of Social Engagement in Child-Robot Interaction}. In
  \bibinfo{booktitle}{\emph{Proceedings of the 9th Nordic Conference on
  Human-Computer Interaction}} (Gothenburg, Sweden)
  \emph{(\bibinfo{series}{NordiCHI '16})}. \bibinfo{publisher}{Association for
  Computing Machinery}, \bibinfo{address}{New York, NY, USA}, Article
  \bibinfo{articleno}{64}, \bibinfo{numpages}{10}~pages.
\newblock
\showISBNx{9781450347631}
\urldef\tempurl%
\url{https://doi.org/10.1145/2971485.2971536}
\showDOI{\tempurl}


\bibitem[\protect\citeauthoryear{Sidner, Lee, Morency, and Forlines}{Sidner
  et~al\mbox{.}}{2006}]%
        {sidner2006effect}
\bibfield{author}{\bibinfo{person}{Candace~L Sidner},
  \bibinfo{person}{Christopher Lee}, \bibinfo{person}{Louis-Philippe Morency},
  {and} \bibinfo{person}{Clifton Forlines}.} \bibinfo{year}{2006}\natexlab{}.
\newblock \showarticletitle{The effect of head-nod recognition in human-robot
  conversation}. In \bibinfo{booktitle}{\emph{Proceedings of the 1st ACM
  SIGCHI/SIGART conference on Human-robot interaction}}.
  \bibinfo{pages}{290--296}.
\newblock


\bibitem[\protect\citeauthoryear{Sinnema and Alimardani}{Sinnema and
  Alimardani}{2019}]%
        {sinnema2019attitude}
\bibfield{author}{\bibinfo{person}{Lizzy Sinnema} {and} \bibinfo{person}{Maryam
  Alimardani}.} \bibinfo{year}{2019}\natexlab{}.
\newblock \showarticletitle{The attitude of elderly and young adults towards a
  humanoid robot as a facilitator for social interaction}. In
  \bibinfo{booktitle}{\emph{Social Robotics: 11th International Conference,
  ICSR 2019, Madrid, Spain, November 26--29, 2019, Proceedings 11}}. Springer,
  \bibinfo{pages}{24--33}.
\newblock


\bibitem[\protect\citeauthoryear{Striepe, Donnermann, Lein, and Lugrin}{Striepe
  et~al\mbox{.}}{2021}]%
        {striepe2021modeling}
\bibfield{author}{\bibinfo{person}{Hendrik Striepe}, \bibinfo{person}{Melissa
  Donnermann}, \bibinfo{person}{Martina Lein}, {and} \bibinfo{person}{Birgit
  Lugrin}.} \bibinfo{year}{2021}\natexlab{}.
\newblock \showarticletitle{Modeling and evaluating emotion, contextual head
  movement and voices for a social robot storyteller}.
\newblock \bibinfo{journal}{\emph{International Journal of Social Robotics}}
  \bibinfo{volume}{13} (\bibinfo{year}{2021}), \bibinfo{pages}{441--457}.
\newblock


\bibitem[\protect\citeauthoryear{Tsoi, Connolly, Ad{\'e}n{\'\i}ran, Hansen,
  Pineda, Adamson, Thompson, Ramnauth, V{\'a}zquez, and Scassellati}{Tsoi
  et~al\mbox{.}}{2021}]%
        {tsoi2021challenges}
\bibfield{author}{\bibinfo{person}{Nathan Tsoi}, \bibinfo{person}{Joe
  Connolly}, \bibinfo{person}{Emmanuel Ad{\'e}n{\'\i}ran},
  \bibinfo{person}{Amanda Hansen}, \bibinfo{person}{Kaitlynn~Taylor Pineda},
  \bibinfo{person}{Timothy Adamson}, \bibinfo{person}{Sydney Thompson},
  \bibinfo{person}{Rebecca Ramnauth}, \bibinfo{person}{Marynel V{\'a}zquez},
  {and} \bibinfo{person}{Brian Scassellati}.} \bibinfo{year}{2021}\natexlab{}.
\newblock \showarticletitle{Challenges deploying robots during a pandemic: An
  effort to fight social isolation among children}. In
  \bibinfo{booktitle}{\emph{Proceedings of the 2021 ACM/IEEE International
  Conference on Human-Robot Interaction}}. \bibinfo{pages}{234--242}.
\newblock


\bibitem[\protect\citeauthoryear{Veletsianos}{Veletsianos}{2012}]%
        {veletsianos2012learners}
\bibfield{author}{\bibinfo{person}{George Veletsianos}.}
  \bibinfo{year}{2012}\natexlab{}.
\newblock \showarticletitle{How do learners respond to pedagogical agents that
  deliver social-oriented non-task messages? Impact on student learning,
  perceptions, and experiences}.
\newblock \bibinfo{journal}{\emph{Computers in Human Behavior}}
  \bibinfo{volume}{28}, \bibinfo{number}{1} (\bibinfo{year}{2012}),
  \bibinfo{pages}{275--283}.
\newblock


\bibitem[\protect\citeauthoryear{Yang}{Yang}{2012}]%
        {yang2012small}
\bibfield{author}{\bibinfo{person}{Wenhui Yang}.}
  \bibinfo{year}{2012}\natexlab{}.
\newblock \showarticletitle{Small talk: A strategic interaction in Chinese
  interpersonal business negotiations}.
\newblock \bibinfo{journal}{\emph{Discourse \& Communication}}
  \bibinfo{volume}{6}, \bibinfo{number}{1} (\bibinfo{year}{2012}),
  \bibinfo{pages}{101--124}.
\newblock


\bibitem[\protect\citeauthoryear{Youssef, Said, Alkork, and Beyrouthy}{Youssef
  et~al\mbox{.}}{2022}]%
        {youssef2022survey}
\bibfield{author}{\bibinfo{person}{Karim Youssef}, \bibinfo{person}{Sherif
  Said}, \bibinfo{person}{Samer Alkork}, {and} \bibinfo{person}{Taha
  Beyrouthy}.} \bibinfo{year}{2022}\natexlab{}.
\newblock \showarticletitle{A survey on recent advances in social robotics}.
\newblock \bibinfo{journal}{\emph{Robotics}} \bibinfo{volume}{11},
  \bibinfo{number}{4} (\bibinfo{year}{2022}), \bibinfo{pages}{75}.
\newblock


\bibitem[\protect\citeauthoryear{Zhou, Mark, Li, and Yang}{Zhou
  et~al\mbox{.}}{2019}]%
        {zhou2019trusting}
\bibfield{author}{\bibinfo{person}{Michelle~X Zhou}, \bibinfo{person}{Gloria
  Mark}, \bibinfo{person}{Jingyi Li}, {and} \bibinfo{person}{Huahai Yang}.}
  \bibinfo{year}{2019}\natexlab{}.
\newblock \showarticletitle{Trusting virtual agents: The effect of
  personality}.
\newblock \bibinfo{journal}{\emph{ACM Transactions on Interactive Intelligent
  Systems (TiiS)}} \bibinfo{volume}{9}, \bibinfo{number}{2-3}
  (\bibinfo{year}{2019}), \bibinfo{pages}{1--36}.
\newblock


\end{thebibliography}

\newpage
\appendix
\section*{Appendix}

\begin{appendix}

\begin{table*}[h]
    
    \centering
    \caption{Overview of Codes and Their Definitions for the Data Analysis: Part A}
    \label{tab:code_overview_a}
    \begin{tabular}{l p{9.5cm}}

    \midrule
    \midrule
    \multicolumn{2}{l}{\textbf{Robot's Verbal Phrases: Task-Oriented}} 
    \\ 
    \midrule
    \midrule
    \textbf{Name} & \textbf{Definition} 
    \\ \hline 
    Confirmation & Robot confirms it processed user request
    \\ \hline
    Reminder & Robot gives a reminder message about the task
    \\

    \midrule
    \midrule
    \multicolumn{2}{l}{\textbf{Robot's Verbal Phrases: Small-Talk}} 
    \\ 
    \midrule
    \midrule
    \textbf{Name} & \textbf{Definition} 
    \\ \hline 
    Question & Robot initiates a question to the user
    \\ \hline
    Statement  & Robot makes a small-talk response or comment
    \\

    \midrule
    \midrule
    \multicolumn{2}{l}{\textbf{Participant's Verbal Phrases: Task-Oriented}} 
    \\ 
    \midrule
    \midrule
    \textbf{Name} & \textbf{Definition} 
    \\ \hline
    Request & User requests a yellow or green pipe, or reports a faulty joint
    \\ \hline
    Reminder & User responses to the robot's reminder message
    \\ \hline
    Confirmation & User responses to the robot's confirmation message
    \\ \hline
    System Announcement & User responses to the negative system announcement message
    \\

        \midrule
    \midrule
    \multicolumn{2}{l}{\textbf{Participant's Verbal Phrases: Small Talk}} 
    \\ 
    \midrule
    \midrule
    \textbf{Name} & \textbf{Definition} 
    \\ \hline
    Question & User initiates a question to the robot
    \\ \hline
    Repeat & User asks the robot to repeat or clarify what it just said
    \\ \hline  
    Short & Non-question small talk statement <= 5 words
    \\ \hline
    Medium  & Non-question small talk statement > 5 words
    \\ \hline
    Long & Multi-sentence non-question small talk statements
    \\
  
    \midrule
    \midrule

    \end{tabular}
    \end{table*}

\begin{table*}[h]
    
    \centering
    \caption{Overview of Codes and Their Definitions for the Data Analysis: Part B}
    \label{tab:code_overview_b}
    \begin{tabular}{l p{10cm}}

    \midrule
    \midrule
    \multicolumn{2}{l}{\textbf{Motives for Initiating Questions}} 
    \\ 
    \midrule
    \midrule
    \textbf{Name} & \textbf{Definition} 
    \\ \hline
    Curiosity	& Individual describes a level of curiosity 
    \\ \hline
    Felt Human-like	& Individual makes connections to human-likeness 
    \\ \hline
    Social Convention & Individual describes factors relating to social conventions
    \\

    \midrule
    \midrule  
    \multicolumn{2}{l}{\textbf{Non-Verbal Reactions to Negative Feedback}} 
    \\ 
    \midrule
    \midrule
    \textbf{Name} & \textbf{Definition} 
    \\ \hline
    Verbal Response	& An audible or verbal phrase spoken by the participant \\ \hline
    Non-Verbal Response & Behaviors such as smiling, laughing, eye-rolls, head nods, etc. \\ 
    
    \midrule
    \midrule
    \multicolumn{2}{l}{\textbf{Blame Attribution}} 
    \\ 
    \midrule
    \midrule
    \textbf{Name} & \textbf{Definition} 
    \\ \hline
    Self & Selects themselves as a potential cause for the feedback announcement
    \\ \hline
    Robot & Selects the robot as a potential cause for the feedback announcement
    \\ \hline
    Team & Selects the team as a potential cause for the feedback announcement
    \\ \hline
    Nobody & Believes neither they nor the robot caused the feedback announcement
    \\

 \midrule
    \midrule
    \multicolumn{2}{l}{\textbf{Response to First Robot Question: Conversation Reciprocation}} 
    \\ 
    \midrule
    \midrule
    \textbf{Name} & \textbf{Definition} 
    \\ \hline
    Answer & Response (Yes) and answered robot's question
    \\ \hline
    Refuse & Response (Yes) but did not answer the robot's question
    \\ \hline
    No engagement	& Participant did not answer the robot's question
    \\
    
    \midrule
    \midrule
    \multicolumn{2}{l}{\textbf{Response to First Robot Question: Impact on Task}} 
    \\ 
    \midrule
    \midrule
    \textbf{Name} & \textbf{Definition} 
    \\ \hline
    Minimal (glance) & Looks at the robot but no task disruption
    \\ \hline
    Freeze & Stops current action (min. 2 sec) and looks at robot before continuing
    \\ \hline
    None & Did not stop or look at the robot
    \\ 
    \midrule
    \midrule
    \multicolumn{2}{l}{\textbf{Response to First Robot Question: Affective Reactions}} 
    \\ 
    \midrule
    \midrule
    \textbf{Name} & \textbf{Definition} 
    \\ \hline
    Pleasant & Individual exhibits a smile
    \\ \hline
    Surprise & Individual exhibits behaviors such as a chin raise, lower lip raise, brow lowering, etc.
    \\ \hline
    Indifference & No reaction 
    \\

    
    \midrule
    \midrule
    \end{tabular}
    \end{table*}

    
  
    


\end{appendix}
\end{document}